\documentclass[PAPER, atlasdraft=false, USenglish, cernpreprint, texmf, orcidlogo]{atlasdoc}
\usepackage[subfigure=false]{atlaspackage}
\usepackage{atlasbiblatex}
\usepackage{multirow}
\usepackage{atlascontribute}

\usepackage{atlasphysics}
\graphicspath{{logos/}{figures/}}

\usepackage{ANA-HION-2023-11-PAPER-defs}
\usepackage{atlasheavyion}
\AtlasTitle{Measurement of forward jet suppression in Pb+Pb collisions at $\sqrt{s_{\mathrm{NN}}}~=~5.02$~\TeV with the ATLAS detector}

\AtlasAbstract{%
Inclusive jet production is measured with the ATLAS detector at the LHC at $\sqrt{s_{\mathrm{NN}}}=5.02$~\TeV in Pb+Pb and $pp$ collisions, using data samples corresponding to integrated luminosities of 1.72~nb$^{-1}$ and 255~pb$^{-1}$, respectively.
The jet nuclear modification factor, $R_{\mathrm{AA}}$, is reported for anti-$k_t$ jets
with radius parameter $R=0.4$ in the forward-rapidity intervals $2.8<|y|<3.2$ and $3.2<|y|<3.6$, probing transverse momenta $p_\mathrm{T}>74$~\GeV. The observed $R_{\mathrm{AA}}$ values show a significant suppression of jet yields in Pb+Pb collisions relative to those measured in $pp$ collisions. The
suppression increases monotonically from peripheral to central collisions and exhibits a mild jet $p_\mathrm{T}$ dependence.
A comparison with previous ATLAS measurements indicates stronger jet suppression in the $2.8<|y|<3.2$ region than in the $|y| < 2.8$ region.
In the $3.2<|y|<3.6$ region, the precision of the measurement does not allow a conclusive comparison with the $|y|<2.8$ region.
These results constitute the first measurement of inclusive jet suppression at forward rapidity in heavy-ion collisions and provide new constraints on flavor, geometric, and medium-density dependence of parton energy loss.
}

\author{The ATLAS Collaboration}

\PreprintIdNumber{CERN-EP-2026-194}

\AtlasJournal{Phys. Lett. B}
\AtlasCoverEgroupAnalysisTeam{atlas-HION-2023-11-analysis-team@cern.ch}

\hypersetup{pdftitle={ATLAS document},pdfauthor={The ATLAS Collaboration}}

\begin{document}
\maketitle

\clearpage

%
%

%

\section{Introduction}
\label{sec:intro}

\newcommand{\paper}{{Letter}\xspace}

Ultrarelativistic heavy-ion collisions at the Relativistic Heavy Ion Collider (RHIC)~\cite{Harrison:2003sb} and the Large Hadron Collider (LHC)~\cite{Evans:2008zzb} allow the creation of a quark--gluon plasma (QGP) -- a state of matter in which quarks and gluons are deconfined from hadrons. A similar form of matter is believed to have existed in the early stages of the evolution of the Universe~\cite{Busza:2018rrf,Shuryak:2014zxa}. Properties of the QGP can be studied using hard processes in which partons (quarks and gluons) with high transverse momentum ($\pt$) are produced via scatterings of incoming partons carrying a large momentum fraction of the nucleon. When high-$\pt$ partons are scattered in elementary collisions, they propagate through the vacuum and their large initial virtuality is reduced by radiating off other partons. This leads to the formation of a parton shower, which is ultimately terminated by hadronization, resulting in free-streaming hadrons propagating as a collimated jet of particles. When high-$\pt$ partons are scattered in heavy-ion collisions, the resulting parton shower traverses the QGP, and its constituents lose energy through both medium-induced gluon radiation and collisional processes, leading to the phenomenon of jet quenching~\cite{Wiedemann:2009sh,Mehtar-Tani:2013pia,Cao:2020wlm}.
Jet quenching has been quantified using a multitude of observables, including inclusive jet yields and nuclear modification factors, as well as modifications of fragmentation functions, jet shapes, substructure observables, dijet observables, and correlations. Jet quenching has also been measured in several types of hard scattering events, including dijet, $\gamma$+jet, and $Z$+jet events~\cite{Connors:2017ptx,Cunqueiro:2021wls,Apolinario:2024equ}.

The typical observable used to quantify jet quenching is the nuclear modification factor, $\RAA$, defined as the ratio of the scaled jet yield in \PbPb collisions to the jet production cross-section in \pp collisions. The \PbPb yield is scaled by the nuclear overlap function, $\TAA$~\cite{Miller:2007ri}, which
represents the per-event nucleon--nucleon luminosity in \PbPb collisions~\cite{HION-2013-06}, and by the total number of minimum-bias events, $N_\mathrm{evt}$, measured in a given centrality interval. The first measurement of jet $\RAA$ revealed a suppression of inclusive jet yield in \PbPb collisions compared to $\pp$ by approximately a factor of two~\cite{HION-2013-06}. This large suppression was later quantified to persist up to a transverse momenta at the \TeV scale~\cite{HION-2017-10} and for jets of various radii~\cite{CMS-HIN-18-014,ALICE:2023waz,HION-2019-09}, as well as for jets initiated by heavy quarks~\cite{CMS-HIN-20-003,ALICE:2024gqr,HION-2018-24}.
By contrast, a larger $\RAA$ was observed for jets produced in the $\gamma$+jet events~\cite{CMS-HIN-16-002,HION-2022-14}, indicating weaker suppression of these jets relative to the inclusive jets.

The differences between the systems discussed above may be related to the initial parton flavor, namely whether the jet is initiated by a quark or a gluon. The fractions of quark- and gluon-initiated jets depend both on the hard process and on the jet kinematics.
The observed suppression is also sensitive to the steepness of the initial parton spectra, because the same energy loss leads to a larger yield suppression for a more steeply falling spectrum.
In mid-rapidity, jets from the $\gamma$+jet system are predominantly initiated by quarks, whereas the fraction of quark-initiated jets in the inclusive sample is significantly smaller. Quark-initiated jets have a narrower parton shower and a smaller overall color charge, both leading to a weaker interaction of the parton shower with the QGP than in the case of gluon-initiated jets~\cite{Spousta:2015fca,Qiu:2019sfj,Zhang:2023oid,Ogrodnik:2024qug}.
The fraction of quark-initiated jets evolves both with $\pt$ and with rapidity, reflecting changes in the parton distribution functions and the relative contributions of the underlying hard-scattering subprocesses~\cite{Ellis:1996mzs,Pablos:2022mrx}.
For inclusive jets, the quark fraction is expected to increase with increasing $\pt$ and rapidity, while the jet $\pt$ spectra become increasingly steep at larger rapidities~\cite{Spousta:2015fca,HION-2022-14}.
The observed differences between inclusive jets and $\gamma$+jets may not only be connected to the initial parton flavor and steepness of the spectrum, but may also be due to a difference in jet selection: in $\gamma$+jet events, selecting on the $\gamma$, for which the QGP is transparent~\cite{ALICE:2024yvg,CMS-HIN-18-016,HION-2012-02}, may lead to a different population of jets than in the inclusive jet system, where both leading-order jets are quenched. This difference in the sampled jet population is commonly referred to as selection bias~\cite{Brewer:2021hmh,Casalderrey-Solana:2016jvj}. All of these considerations motivate measurements of jet suppression across kinematic regions with different expected jet flavor compositions.

This \paper reports the first measurement at forward rapidity of inclusive jet yields in heavy-ion collisions and inclusive jet cross-sections in \pp collisions, both at $\snn$=5.02 \TeV, using data recorded with the ATLAS detector at the LHC.
Jets are reconstructed using the \antikt\ algorithm~\cite{Cacciari:2008gp} with radius parameter $R=0.4$ and measured for $\pt>74$~\GeV in two forward rapidity intervals, \semifwd and \fwd.
Forward jet suppression is quantified using the jet $\RAA$, measured in four centrality intervals between 0\% and 80\%. While many measurements of inclusive jet suppression were performed in the past, all of them were done in the central region ($|y| \lesssim 3$), where gluon-initiated jets dominate at $\pt \lesssim 100$~\GeV~\cite{Cunqueiro:2021wls}.
Measurements of jet suppression at forward rapidity therefore provide an important complement to studies at mid-rapidity, helping to disentangle the flavor dependence of jet quenching, selection biases~\cite{Pablos:2022mrx}, and other factors affecting parton energy loss, including its path-length dependence in an expanding QGP~\cite{Adhya:2021kws,Adhya:2024pgm}.

%


\section{ATLAS detector}
\label{sec:detector}

\newcommand{\AtlasCoordFootnote}{
ATLAS uses a right-handed coordinate system with its origin at the nominal interaction point (IP)
in the center of the detector and the \(z\)-axis along the beam pipe.
The \(x\)-axis points from the IP to the center of the LHC ring,
and the \(y\)-axis points upwards.
Cylindrical coordinates \((r,\phi)\) are used in the transverse plane,
\(\phi\) being the azimuthal angle around the \(z\)-axis.
The pseudorapidity is defined in terms of the polar angle \(\theta\) as \(\eta = -\ln \tan(\theta/2)\).
The rapidity is defined as \(y\) = \(0.5 \ln[(E+p_{z})/(E-p_{z})]\) where \(E\) and \(p_{z}\) are the energy and \(z\)-component of the momentum along the beam direction, respectively.
Angular distance is measured in units of \(\Delta R \equiv \sqrt{(\Delta\eta)^{2} + (\Delta\phi)^{2}}\).}

The ATLAS detector~\cite{PERF-2007-01} at the LHC is a multipurpose particle detector with a forward--backward symmetric cylindrical geometry and a near-\(4\pi\) coverage in solid angle.\footnote{\AtlasCoordFootnote}
It consists of an inner tracking detector surrounded by a thin superconducting solenoid, electromagnetic and hadronic calorimeters, and a muon spectrometer.
The inner-detector system is immersed in a \SI{2}{\tesla} axial magnetic field and provides charged-particle tracking within \(|\eta| < 2.5\).
The high-granularity silicon pixel detector covers the vertex region and typically provides four measurements per track, with the first hit typically being in the insertable B-layer installed before Run~2~\cite{ATLAS-TDR-2010-19,PIX-2018-001}.
It is followed by the silicon microstrip tracker (SCT) that typically provides eight measurements per track.
These silicon detectors are complemented by the transition radiation tracker (TRT), based on longitudinal drift tubes, which surrounds the SCT and has coverage up to \(|\eta| = 2.0\).

The calorimeter system covers the pseudorapidity range \(|\eta| < 4.9\).
Within the region \(|\eta|< 3.2\), electromagnetic calorimetry is provided by barrel and
endcap high-granularity lead/liquid-argon (LAr) calorimeters,
with an additional thin LAr presampler covering \(|\eta| < 1.8\)
to correct for energy loss in the material upstream of the calorimeters.
Hadronic calorimetry is provided by the steel/scintillator-tile calorimeter,
segmented into three barrel structures within \(|\eta| < 1.7\), and two copper/LAr hadronic endcap calorimeters.
The solid angle coverage is completed with copper/LAr and tungsten/LAr calorimeter modules (FCal),
covering the forward regions of $3.1 < |\eta| < 4.9$.
The zero-degree calorimeters (ZDC) consist of layers of alternating quartz rods and tungsten plates which are located symmetrically at $z=\pm140$~m and cover $|\eta| \geq 8.3$. In \PbPb\ collisions, the ZDCs primarily measure ``spectator'' neutrons that do not interact with the incoming nucleus~\cite{Avoni:2025lmu}.
The luminosity is measured mainly by the LUCID--2~\cite{LUCID2} detector that is located close to the beampipe.

The events of interest were selected by the first-level (L1) trigger system implemented in custom hardware, followed by selections made by algorithms implemented in software in the high-level trigger (HLT)~\cite{TRIG-2019-04,ATL-DAQ-PUB-2018-002,ATL-DAQ-PUB-2019-001}.
The L1 jet triggers are based on localized transverse-energy deposits reconstructed from coarse-granularity calorimeter information. The L1 jet triggers used in this analysis require these deposits to exceed transverse-energy thresholds ranging from 12 to 30~\GeV at the electromagnetic energy scale for both central jet triggers, $|\eta| < 3.2$, and forward jet triggers, $3.2 < |\eta| < 4.9$, in both \pp and \PbPb collisions.
The HLT jet trigger employs a jet reconstruction algorithm similar to that used in the offline analysis, which is described in Section~\ref{sec:jetreco}. In \PbPb collisions, three central jet triggers with HLT transverse-energy thresholds ranging from 50 to 75~\GeV are used, while the forward jet trigger uses a threshold of 50~\GeV. In \pp collisions, four central jet triggers with HLT thresholds ranging from 45 to 110~\GeV are used, while two forward jet triggers use thresholds of 45 and 60~\GeV.
In \pp collisions, the highest-threshold jet trigger in each region is unprescaled, while jet triggers with lower thresholds are prescaled and therefore record only a fraction of the events satisfying their trigger conditions.
The $\pt$ thresholds are set such that each trigger is fully efficient within the corresponding jet $\pt$ interval used in the measurement.
In addition to the jet trigger, three triggers are used in \PbPb\ collisions to select minimum-bias events. These minimum-bias triggers used conditions on the transverse energy recorded in the whole calorimeter system by L1 trigger, or a coincidence from both arms of the ZDC trigger, associated with a track reconstructed by the HLT.
An extensive software suite~\cite{ATL-SOFT-PUB-2021-001} is used in data simulation, in the reconstruction and analysis of real and simulated data, in detector operations, and in the trigger and data acquisition systems of the experiment.


\section{Data and simulated event samples}
\label{sec:datamc}

The \PbPb data used in these measurements were collected in 2018, and the \pp\ data used were collected in 2017, both at a per-nucleon-pair center-of-mass energy $\sqrt{s_\mathrm{NN}} = 5.02~\TeV$. Events were selected by jet triggers described in Section~\ref{sec:detector}.
Both \PbPb and \pp data contain signal from multiple inelastic interactions per bunch crossing, also referred to as `pileup'. The average number of inelastic interactions per bunch crossing reaches a maximum value of 0.004 in \PbPb collisions, and it varies between 1.4 and 4.4 in \pp collisions.
Although only a small fraction of the \PbPb events ($<0.5$\%) contain multiple collisions, these are removed by rejecting events that deviate significantly from the expected anti-correlation between the total transverse energy deposited in both forward calorimeters, $\sumETFCal$, and the ZDC energy, which is driven by the nuclear geometry, as well as from the expected correlation between $\sumETFCal$ and the number of reconstructed tracks~\cite{HION-2021-09}.

Pileup collisions are not rejected in the \pp data sample.

The degree of overlap of the two colliding nuclei in \PbPb collisions is characterized by the event centrality, which is estimated from $\sumETFCal$. This estimate is obtained by
associating different ranges of $\sumETFCal$ measured in minimum-bias collisions with an underlying \PbPb collision geometry according to a Monte Carlo (MC) Glauber simulation~\cite{Miller:2007ri,HION-2016-07, HION-2018-06}.
The centrality intervals considered in this measurement are 0--10\%, 10--20\%, 20--50\%, and 50--80\%, ordered from the largest $\sumETFCal$ values, corresponding to the most central collisions, to the smallest $\sumETFCal$ values, corresponding to peripheral collisions.

To correct for the contribution of jets to the centrality estimator, the transverse energy of jets with $\pt>40$~\GeV is subtracted from $\sumETFCal$. This subtraction has only a minor impact on the centrality determination, as the migration of events between adjacent centrality intervals due to the presence of jets is found to be less than 3\%.
The values of the mean nuclear overlap function, \avgTAA~\cite{Miller:2007ri}, and their uncertainties~\cite{HION-2018-25} determined by MC Glauber simulation are listed in Table~\ref{tab:taa}. %

\begin{table}[htbp]
\centering
\small
\caption{
Centrality intervals and corresponding values of mean nuclear overlap function, $\avgTAA$, with systematic uncertainties.
}
\begin{tabular}{|l|c|c|}
\hline
Centrality & $\avgTAA$ [mb$^{-1}$] \\
\hline
0--10\%   &  23.21  $\pm$ 0.06  \\
10--20\%  &  14.38    $\pm$ 0.12  \\
20--50\%  &   5.53    $\pm$ 0.11  \\
50--80\%  &  0.73    $\pm$ 0.04   \\
\hline
\end{tabular}
\label{tab:taa}
\end{table}

The MC simulations used in this analysis are multijet events modeled with the \PYTHIA~{8}~\cite{Sjostrand:2014zea} MC generator with leading-order matrix elements for dijet production, which were matched to the parton shower. The A14 set of tuned parameters~\cite{ATL-PHYS-PUB-2014-021} and the \NNPDF[23lo] parton distribution functions~\cite{Ball:2012cx} were used. The events generated by \PYTHIA[8] were passed through a \GEANT-based simulation~\cite{Agostinelli:2002hh,SOFT-2010-01} of the ATLAS detector. %
In \PbPb collisions, a large underlying event (UE) is produced by multiple soft interactions between nucleons of colliding nuclei.
Therefore, for \PbPb collisions, the simulated dijet events were overlaid with minimum bias \PbPb data, and this `overlay' sample was reweighted on an event-by-event basis to obtain the same $\sumETFCal$ distribution
as in the jet-triggered \PbPb data sample.
This overlay procedure provides a realistic description of the UE contribution to jet reconstruction and performance, since the simulated hard-scattering events are embedded into real minimum-bias \PbPb data events.
The MC events are reconstructed in the same way as data events.


\section{Jet reconstruction and analysis procedure}
\label{sec:jetreco}

The jet reconstruction procedures follow those used by the ATLAS Collaboration for previous jet measurements in \PbPb\ collisions~\cite{HION-2017-10,HION-2020-09}.
Jets are reconstructed using the \antikt\ algorithm implemented in the FastJet software package~\cite{Fastjet}. In both \pp\ and \PbPb\ collisions, jets with $R = 0.4$ are formed by clustering calorimeter towers with $\Delta\eta \times\Delta\phi=0.1\times\pi/32$ in the pseudorapidity range $|\eta|< 4.9$ over the full azimuthal range.
The energies in the towers are obtained by summing the energies of calorimeter cells at the electromagnetic energy scale~\cite{PERF-2011-03} within the tower boundaries. The UE contribution to each tower is subtracted on an event-by-event basis by estimating the average local energy density, $\rho(\eta,\phi)$. This estimate is obtained by measuring the average energy in calorimeter towers as a function of $\eta$ and including the azimuthal modulation due to harmonic flow characterized by the second-, third-, and fourth-order flow coefficients~\cite{HION-2016-06}.
Following the UE subtraction, new estimates of $\rho$ are obtained by excluding the contributions coming from towers within jets from the average energy density and flow parameterization.
Additionally, the UE is also corrected for $\eta$- and $\phi$-dependent non-uniformities in the detector response to soft particles by applying correction factors derived in minimum-bias \PbPb\ data. In \pp\ collisions, the same background subtraction procedure is used, but without the non-uniformity correction and harmonic modulation of the UE.

Following the UE subtraction, $\eta$- and $\pT$-dependent multiplicative calibration factors derived in MC simulations are applied to the jet four-momentum vector to account for the non-compensating hadronic response of the calorimeter~\cite{ATLAS-CONF-2015-016}.
An additional correction based on \textit{in situ} studies of jets recoiling against photons, \Zboson\ bosons, and jets in other regions of the calorimeter is applied to account for differences between data and the simulation~\cite{HION-2018-06}. This calibration is followed by a cross-calibration which relates the jet energy scale (JES) of jets reconstructed using the procedure outlined in this section to the JES in 13~\TeV\ \pp\ collisions that use a different jet reconstruction technique~\cite{ATLAS-CONF-2015-016}.

Generator-level jets, also referred to as truth jets, %
are defined in the MC samples before detector simulation by applying the \antikt\ algorithm with $R=0.4$ to stable particles with a proper lifetime greater than 30~ps, but excluding muons and neutrinos, which do not leave significant energy deposits in the calorimeter.
After the detector simulation, the truth jets are matched to the nearest reconstructed jet with $\Delta R <0.3$.
The performance of the jet reconstruction is characterized by the JES and jet energy resolution (JER), which correspond to the mean and variance, respectively, of the $\pt^{\mbox{\tiny{reco}}}/\pt^{\mbox{\tiny{truth}}}$ distribution, where $\pt^{\mbox{\tiny{reco}}}$ is the reconstructed jet \pt\ and $\pt^{\mbox{\tiny{truth}}}$ is the \pt of the matched truth jet. The deviation of JES from unity is centrality independent and smaller than 4\%. The impact of this small deviation on measured quantities is corrected by the unfolding procedure described below.
\begin{figure}
\begin{subfigure}{0.45\textwidth}
\includegraphics[width=\linewidth]{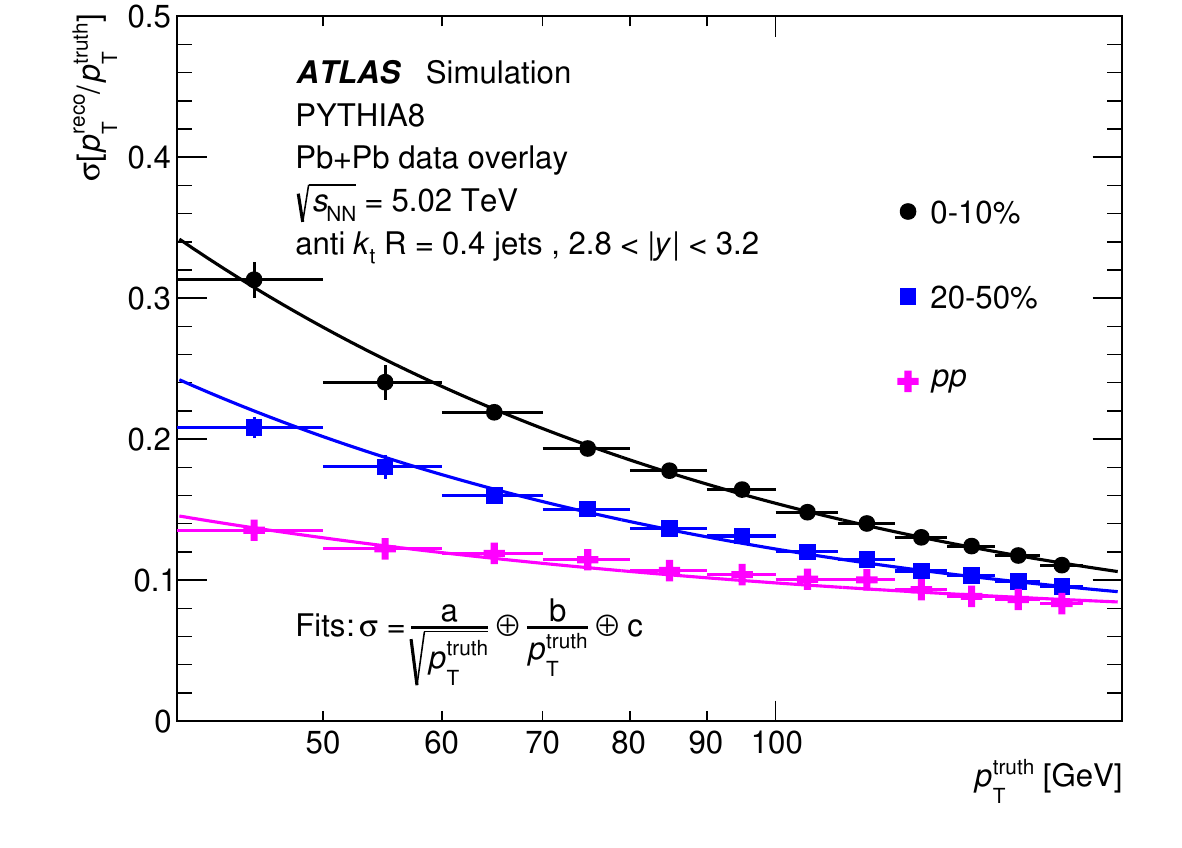}
\caption{} %
\end{subfigure}
\begin{subfigure}{0.45\textwidth}
\includegraphics[width=\linewidth]{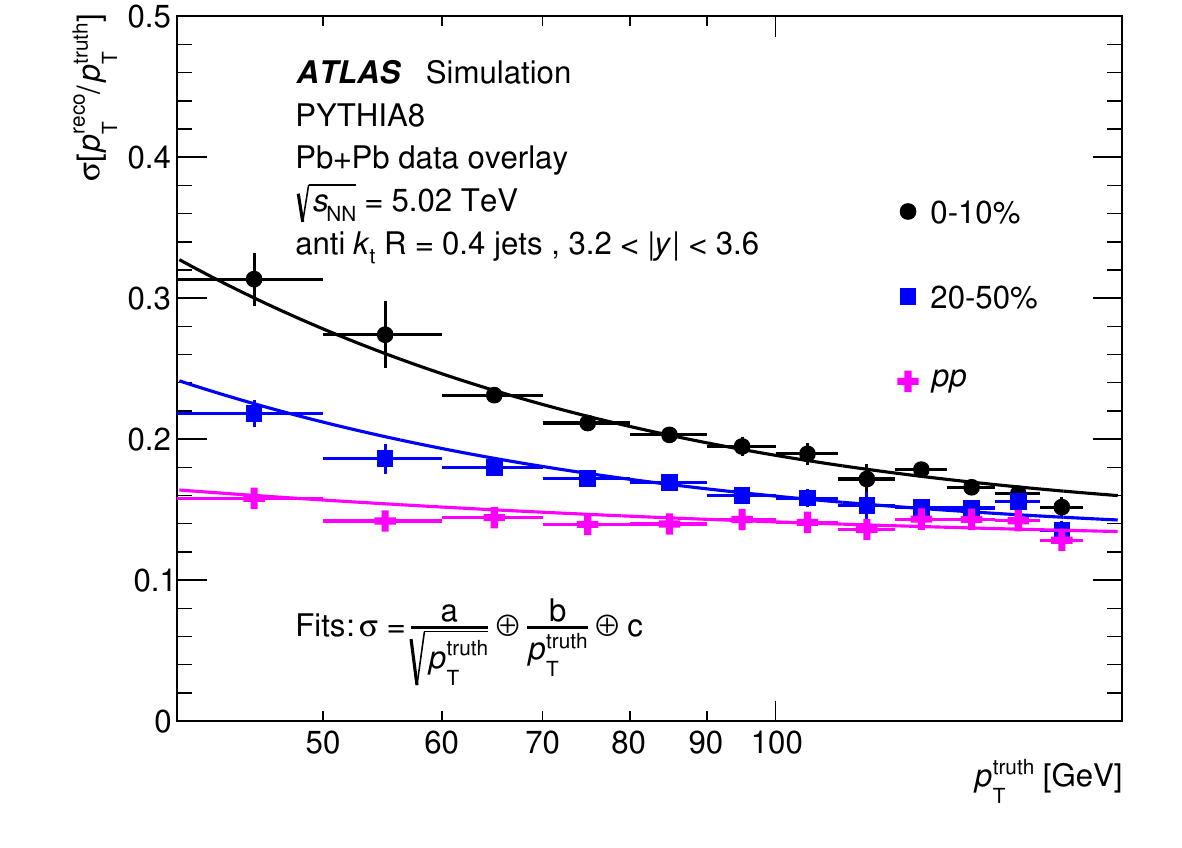}
\caption{} %
\end{subfigure}
\caption{The jet energy resolution evaluated as a function of truth jet $\pt$ for (a) \semifwd and (b) \fwd rapidity regions. The JER is evaluated for 0--10\%, 20--50\% centrality in \PbPb and in $\pp$ simulation.}
\label{fig:jer}
\end{figure}

The JER is shown as a function of $\pt^{\mbox{\tiny{truth}}}$ in Figure~\ref{fig:jer}. The JER is evaluated separately for truth jets in the \semifwd and \fwd regions, and for 0--10\% and 20--50\% \PbPb and \pp simulations.
To express the different contributions, the JER is parameterized by a quadrature sum of three terms,
\begin{equation}
\sigma\bigg(
\frac{p_\mathrm{T}^{\mbox{\tiny{reco}}}}{p_\mathrm{T}^{\mbox{\tiny{truth}}}}
\bigg) = \frac{a}{\sqrt{ p_\mathrm{T}^{\mbox{\tiny{truth}}}}}\oplus\frac{b}{ p_\mathrm{T}^{\mbox{\tiny{truth}}}}\oplus{c}.
\label{eq:jer}
\end{equation}
The three terms are the stochastic, noise, and constant terms, quantified by the parameters $a$, $b$, and $c$, respectively. The parameters $a$ and $c$ characterize contributions related to the detector response and were found to be independent of centrality~\cite{HION-2017-10}. The parameter $b$ is centrality dependent and, in \PbPb collisions, is driven by UE fluctuations uncorrelated with the jet $\pt$.
Although the JER parameterization in Equation~(\ref{eq:jer}) is not used in any of the corrections in this \paper, it provides a quantitative characterization of the impact of detector effects and UE deposits on the performance of jet reconstruction in the forward region. It may therefore serve as useful input for future studies of forward jets in heavy-ion collisions.

To stabilize the fit of the JER %
across systems and rapidity intervals, the value of parameter $a$ was determined from fitting the $\pp$ simulation in the rapidity region of $2.8<|y|<3.2$. This yields a value of $a=0.68 \pm 0.01 \sqrt{\text{\GeV}}$, %
which is then used in all other fits.
Table~\ref{tab:jer} summarizes the values of the $b$ and $c$ parameters for truth jets in the \semifwd and \fwd regions.
The $b$ term is significantly smaller in the \fwd region than in the \semifwd region, which is consistent with smaller overall UE deposits with increasing rapidity.
The $c$ term is significantly larger in the \fwd region than in the \semifwd region, which is consistent with a less uniform calorimeter response and larger material-related effects in the forward region.

\begin{table}
\centering
\caption{Values of noise term ($b$) and constant term ($c$) obtained from the fit of jet energy resolution in \semifwd and \fwd rapidity regions and for $\pp$ and \PbPb collisions. The quoted uncertainties are statistical only.}
\begin{tabular}{|c|c|c|c|}
\hline
Rapidity & System & $b [\mathrm{\GeV}]$ & $c$ \\
\hline
\multirow{5}{*}{\semifwd}
& \PbPb, 0--10\% & $14.57 \pm 0.15$ & $0.049 \pm 0.004$ \\
\cline{2-4}
& \PbPb, 10--20\% & $12.35 \pm 0.09$ & $0.050 \pm 0.003$ \\
\cline{2-4}
& \PbPb, 20--50\% & $8.89 \pm 0.07$ & $0.054 \pm 0.002$ \\
\cline{2-4}
& \PbPb, 50--80\% & $4.81 \pm 0.10$ & $0.058 \pm 0.002$ \\
\cline{2-4}
& $\pp$ & $2.98 \pm 0.06$ & $0.060 \pm 0.001$ \\
\hline
\multirow{5}{*}{\fwd}
& \PbPb, 0--10\% & $11.43 \pm 0.28$ & $0.124 \pm 0.004$ \\
\cline{2-4}
& \PbPb, 10--20\% & $8.49 \pm 0.28$ & $0.132 \pm 0.003$ \\
\cline{2-4}
& \PbPb, 20--50\% & $7.05 \pm 0.24$ & $0.126 \pm 0.002$ \\
\cline{2-4}
& \PbPb, 50--80\% & $4.24 \pm 0.45$ & $0.124 \pm 0.002$ \\
\cline{2-4}
& $\pp$ & $1.65 \pm 0.07$ & $0.122 \pm 0.001$ \\
\hline
\end{tabular}
\label{tab:jer}
\end{table}

A minimum \pt threshold of 74~\GeV is chosen in this measurement to ensure a fully efficient jet reconstruction at the trigger level and offline.
The threshold further ensures that the contribution of jets reconstructed only from fluctuations of the underlying events (``UE jets'')
is reduced to negligible level in the \PbPb\ measurement.
The method typically used for the removal of UE jets in Pb+Pb measurements relies on the tracking information~\cite{HION-2017-10}, which is not available in the forward region.
Instead, the contribution from UE jets is estimated by fitting jets \pt distribution by a power law, describing the hard-process contribution~\cite{Spousta:2015fca}, and a Gaussian distribution, describing the UE jet contribution at low \pt~\cite{ALICE:2012nbx}.
The values of the Gaussian width, obtained from the fit are cross-checked independently by a fluctuation study following~Ref.~\cite{HION-2011-02}. In this study, the distribution of the sums of transverse energies in calorimeter towers, $\sum E_\mathrm{T}^\mathrm{twr}$, is evaluated in minimum bias events. The $\sum E_\mathrm{T}^\mathrm{twr}$ are calculated in the $\Delta\eta \times \Delta\phi$ regions of size $0.7 \times 0.7$, which corresponds to an average area of an \antikt $R=0.4$ jet. The widths of $\sum E_\mathrm{T}^\mathrm{twr}$ distributions are found to be consistent with those obtained from fits.
The contamination coming from UE jets does not exceed 0.1\% in any reported \pT interval of any distribution measured in \PbPb\ collisions.

The jet \pt spectra are unfolded using a one-dimensional Bayesian unfolding method~\cite{DAgostini:1994fjx} from the \verb=RooUnfold= software package~\cite{Adye:2011gm}. This accounts for resolution effects, including those from migration between adjacent rapidity intervals. The response matrices used in the unfolding are built from pairs of truth jets matched with reconstructed jets in the simulation.
The response matrices are generated separately in \pp\ and each of the measured centrality intervals in \PbPb collisions. The response matrices are reweighted along the truth-jet \pt axis by the data-to-MC ratio of reconstructed jet \pt spectra in order to obtain a prior which better matches observation. The number of iterations used in the unfolding is determined by minimizing the quadrature sum of the total statistical uncertainty and the differences in the unfolded distribution between consecutive iterations. This number is set to four for all selections used in the measurement. When tested in simulation, this unfolding procedure recovers the original generator-level distribution within the statistical uncertainties of the test sample.


\section{Systematic uncertainties}
\label{sec:syst}

The systematic uncertainties considered in this measurement originate from the JES, JER, the unfolding procedure, the determination of the mean nuclear overlap function $\avgTAA$, and the integrated luminosity of the \pp\ reference data sample. They are separated into two classes. The JES, JER, and unfolding uncertainties modify the shapes of the measured spectra and are treated as bin-wise correlated
uncertainties. The uncertainties in $\avgTAA$ and in the \pp\ luminosity affect only the overall normalization of the corresponding distributions and of the \RAA\ measurement.

For each source of uncertainty that affects the reconstructed jet response, the analysis is repeated using a modified response matrix, and the difference between the resulting unfolded distribution and the nominal result is assigned as the corresponding systematic uncertainty. The individual contributions are evaluated separately and combined in quadrature.

\begin{figure}
\centering
\includegraphics[width=0.327\linewidth]{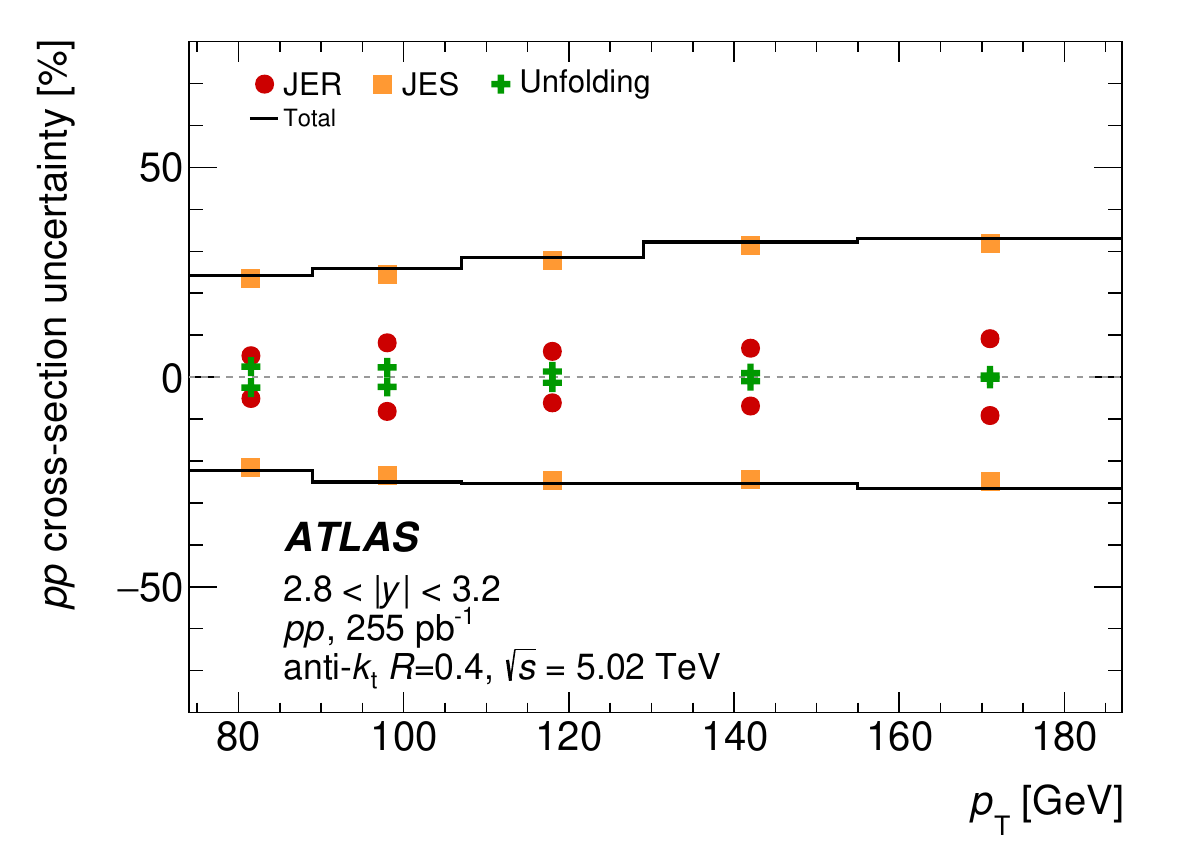}%
\hfill
\includegraphics[width=0.327\linewidth]{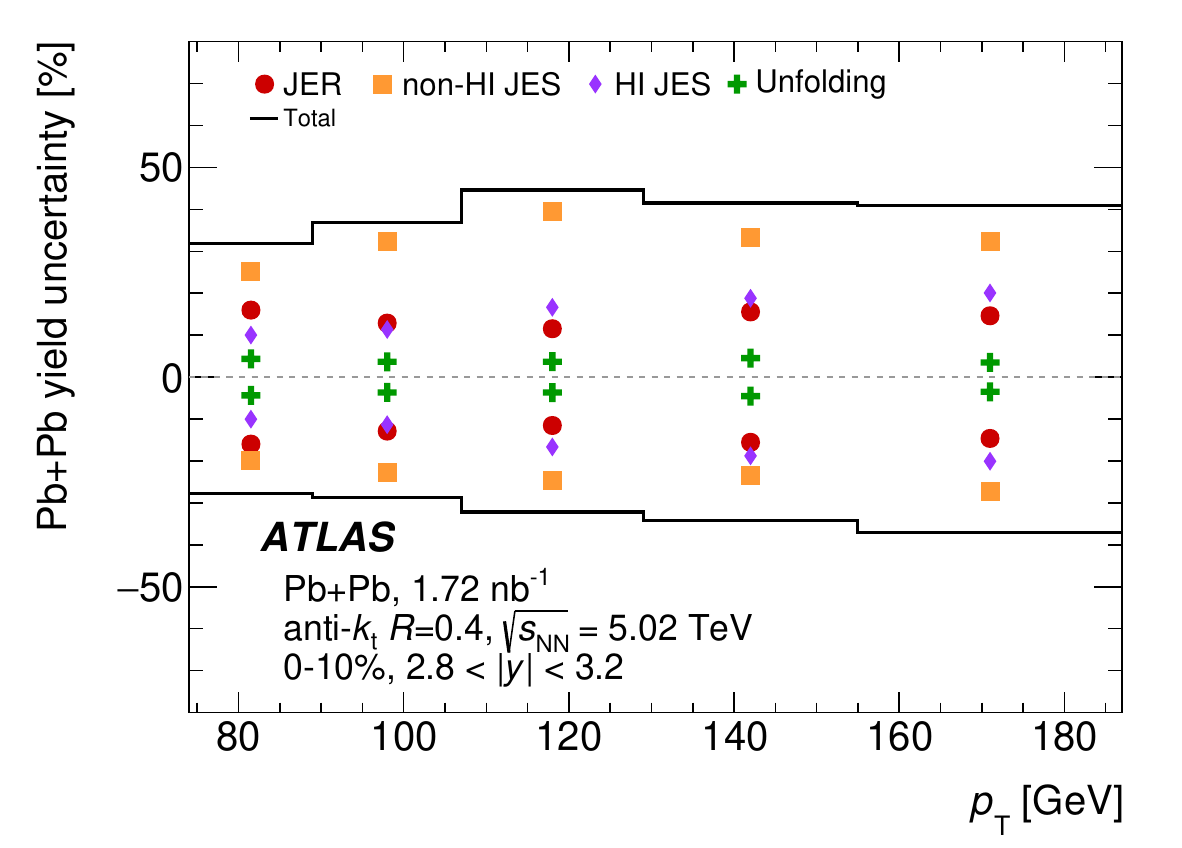}%
\hfill
\includegraphics[width=0.327\linewidth]{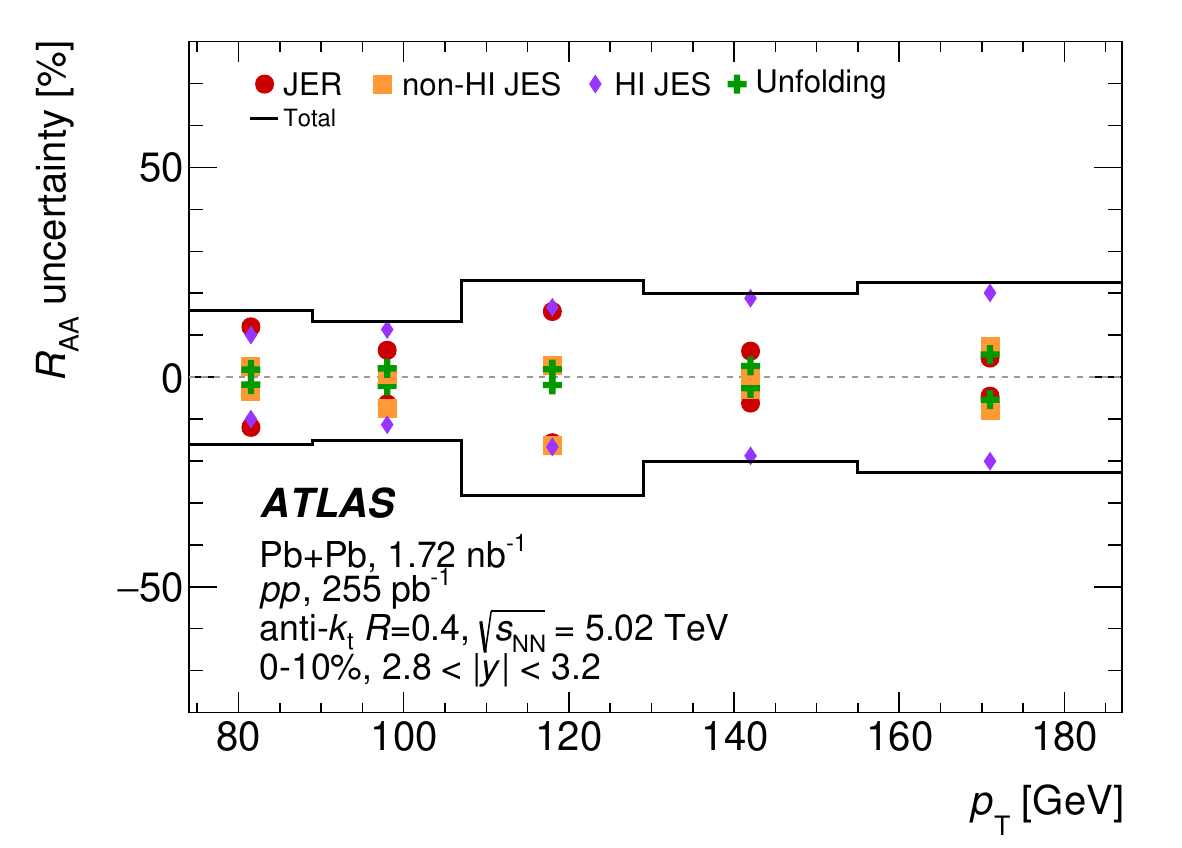}%
\\
\includegraphics[width=0.327\linewidth]{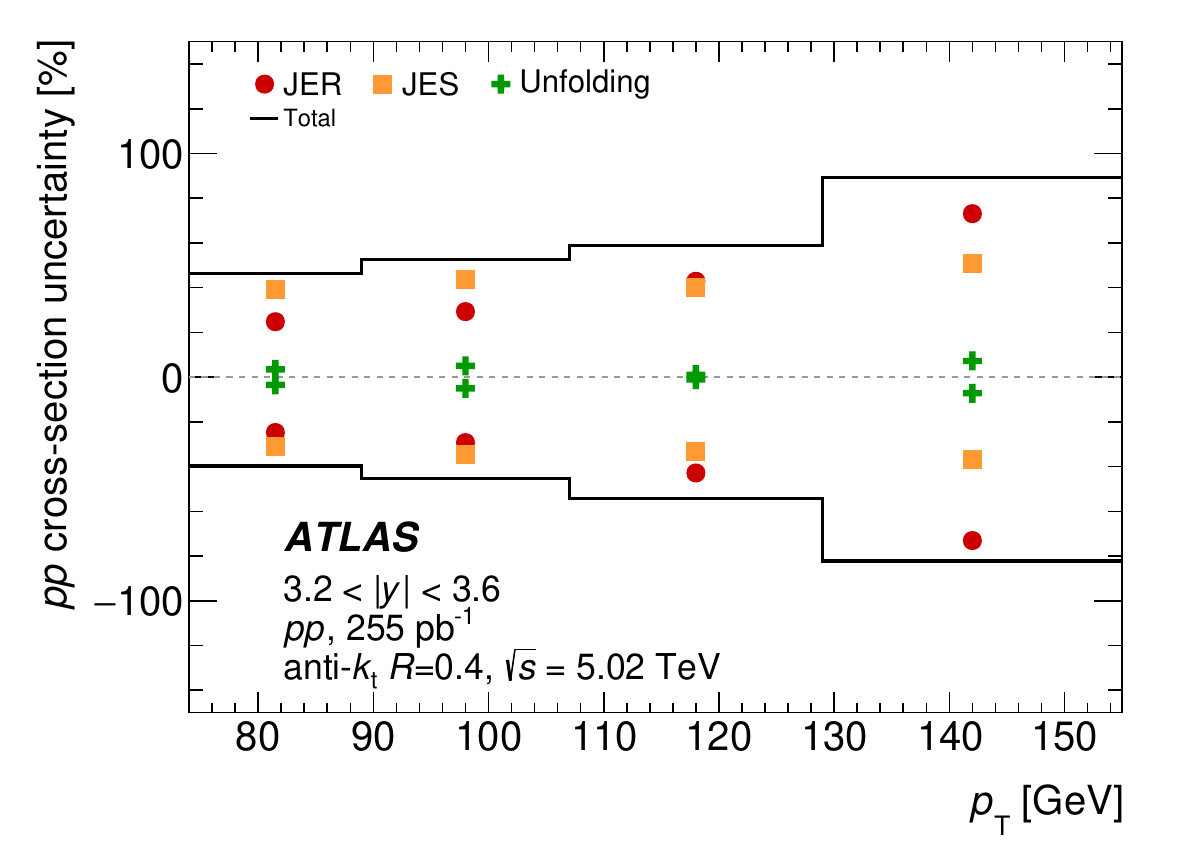}%
\hfill
\includegraphics[width=0.327\linewidth]{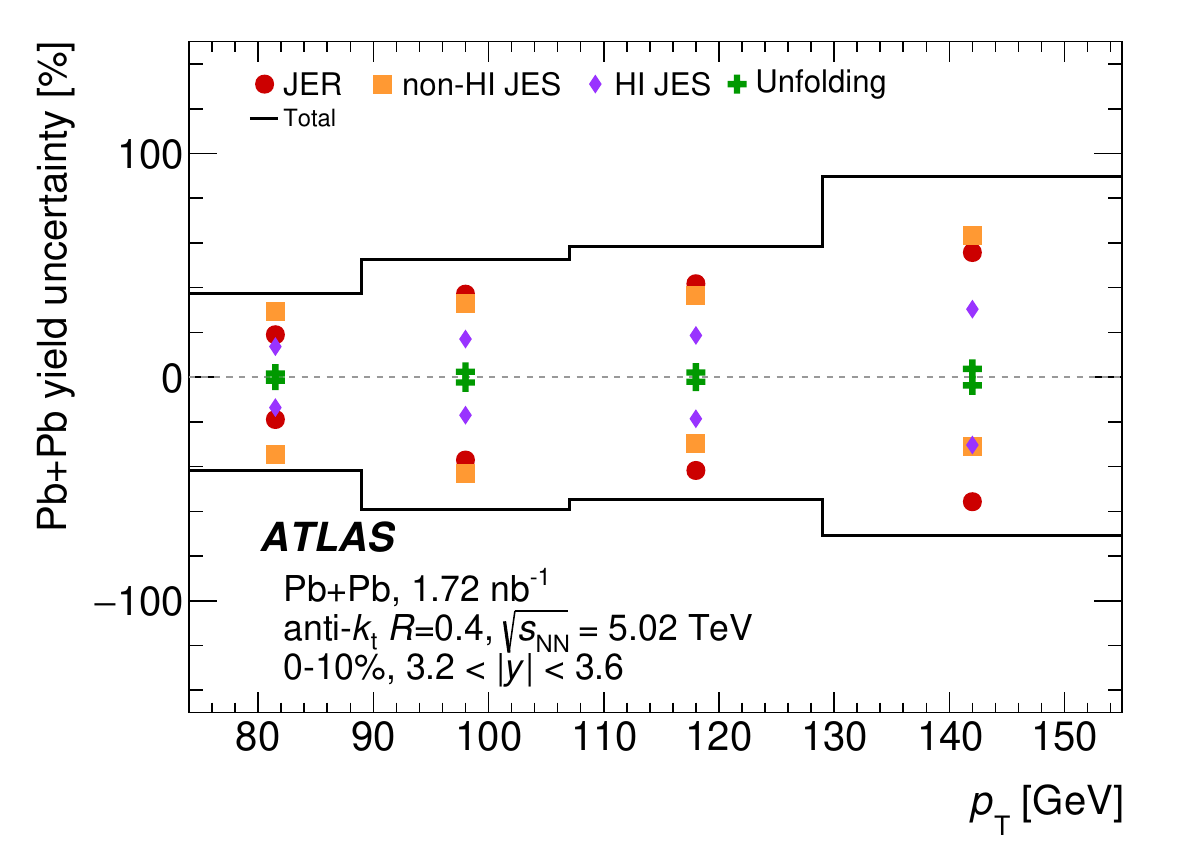}%
\hfill
\includegraphics[width=0.327\linewidth]{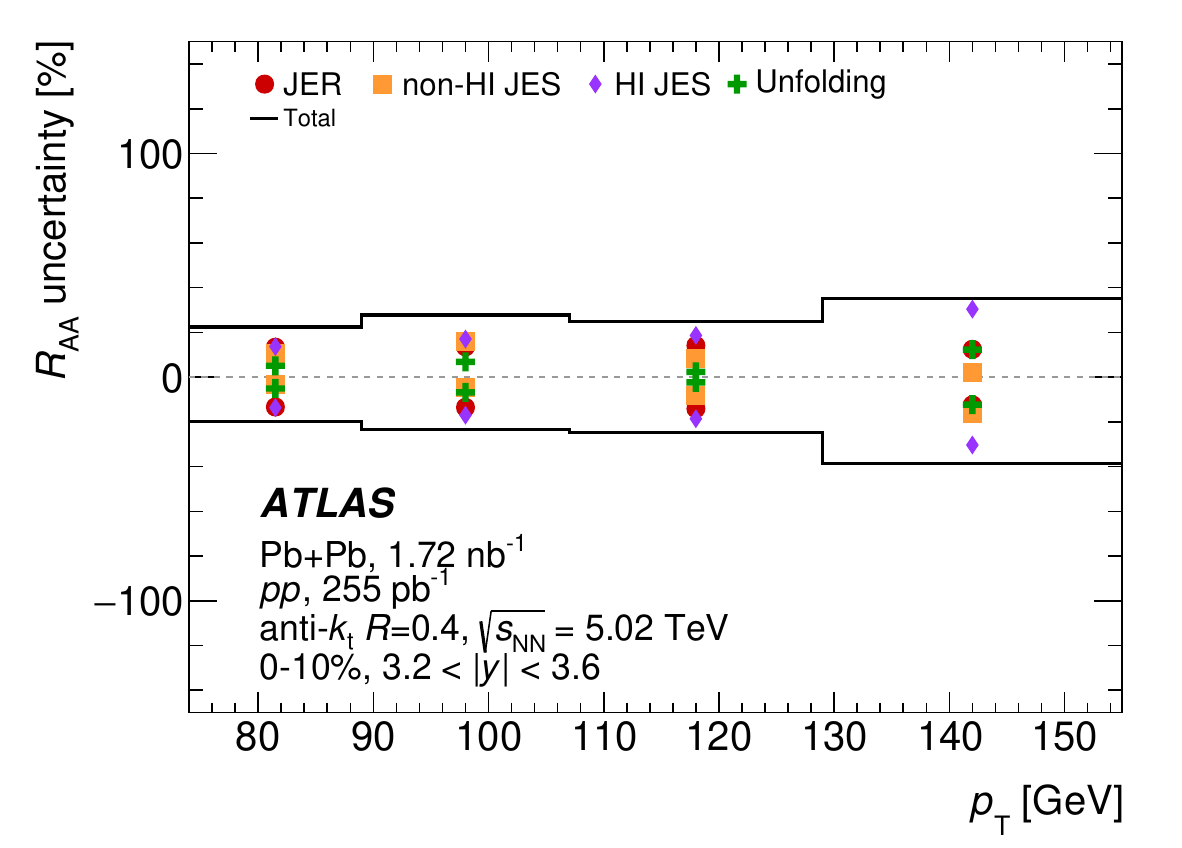}%
\\
\caption{Systematic uncertainties, for \pp\ jet cross-sections (left), \PbPb\ jet yields (middle) and jet \RAA\ (right). Upper and lower panels show uncertainties for rapidity selection of \semifwd and \fwd, respectively. Systematic uncertainties on \pp\ luminosity and \avgTAA, which affect the overall normalization of measured distributions, are not shown. }
\label{fig:sys}
\end{figure}

The JES uncertainty contains components that are common to \pp\ and \PbPb\ collisions, as well as a component specific to the heavy-ion environment. In the uncertainty breakdown, shown in Figure~\ref{fig:sys}, the common component is denoted as JES for the \pp\ result and as `non-HI JES' for the \PbPb\ result, while the additional component specific to the \PbPb\ environment is denoted as `HI JES'. The centrality-independent part used in \PbPb collisions, and that used for \pp collisions, is derived from \textit{in situ} studies of the calorimeter response to jets in \pp\ collisions~\cite{JETM-2018-05}, together with uncertainties associated with the relative energy scale between the jet reconstruction and calibration procedures used in heavy-ion and \pp\ analyses~\cite{ATLAS-CONF-2015-016}.
A further JES component accounts for possible modifications of the detector response to jets reconstructed in the \PbPb\ environment. Such a modification can result from quenching-induced changes of the jet fragmentation pattern, which are not described by the MC simulation~\cite{HION-2017-04,CMS-HIN-16-020}.
This contribution is evaluated with the method described in Ref.~\cite{ATLAS-CONF-2015-016}. The jet transverse momentum measured in the calorimeter is compared with the scalar sum of the transverse momenta of charged particles associated with the jet in the rapidity region of $|y|<2.1$. The charged particles used in this comparison are required to satisfy $\pt > 4~\GeV$ in order to suppress contributions from the underlying event. The data-to-MC ratio of this quantity is compared between \pp\ and \PbPb\ collisions, and the observed centrality-dependent difference is used to estimate the heavy-ion-specific JES uncertainty.
The same heavy-ion-specific JES uncertainty is applied in the forward rapidity region, since no significant rapidity dependence of jet fragmentation was observed in previous measurements~\cite{HION-2017-04}.
Each JES component is applied as a positive and negative variation of the reconstructed jet \pt\ in simulation, as a function of jet kinematics and, where relevant, collision centrality. %

The uncertainty associated with the JER is evaluated in an analogous way. The reconstructed jet transverse momentum in the MC samples is smeared by an additional Gaussian contribution, and the response matrices are regenerated. The smearing factors are derived from \textit{in situ} measurements of the jet resolution in \pp\ collisions, based on dijet transverse-momentum or energy balance studies~\cite{JETM-2018-05}. An additional uncertainty is included to account for differences between the heavy-ion jet reconstruction procedure and that used in the corresponding 13~\TeV \pp\ calibration studies. The unfolded distributions obtained with the modified JER response matrices are compared with the nominal results, and the resulting deviations are symmetrized.

The sensitivity of the result to the unfolding procedure is evaluated by varying the construction of the response matrix. In the nominal analysis, the response matrix is reweighted along the truth jet axis in order to improve the description of the measured jet spectra. The unfolding uncertainty is estimated from the difference between the nominal result and the result obtained using response matrices constructed without this reweighting. This variation probes the dependence of the unfolded spectra on the assumed prior used in the unfolding. The resulting deviations are symmetrized.

The uncertainty in $\avgTAA$ arises from the Glauber-model determination of the collision geometry. It includes contributions from the uncertainty in the inelastic nucleon--nucleon cross-section, the Woods--Saxon parameterization of the nuclear density distribution, and the fraction of selected inelastic \PbPb\ collisions~\cite{Miller:2007ri}. The corresponding values are shown in Table~\ref{tab:taa}. These uncertainties are treated as normalization uncertainties that are correlated across jet \pt\ intervals within a given centrality interval. The uncertainty in the integrated luminosity of the \pp\ data sample is 1.0\%, derived using methods described in Ref.~\cite{DAPR-2021-01}. The luminosity uncertainty is also treated as a global normalization uncertainty. %

The magnitude of bin-wise correlated systematic uncertainties is summarized in Figure~\ref{fig:sys} separately for jet cross-sections in \pp\ collisions, jet yields in the 0--10\% centrality interval of \PbPb collisions, and for the $\RAA$ measured in the same centrality interval.

\newpage

\section{Results}
\label{sec:result}
\FloatBarrier
\begin{figure}
\centering
\includegraphics[width=0.45\linewidth]{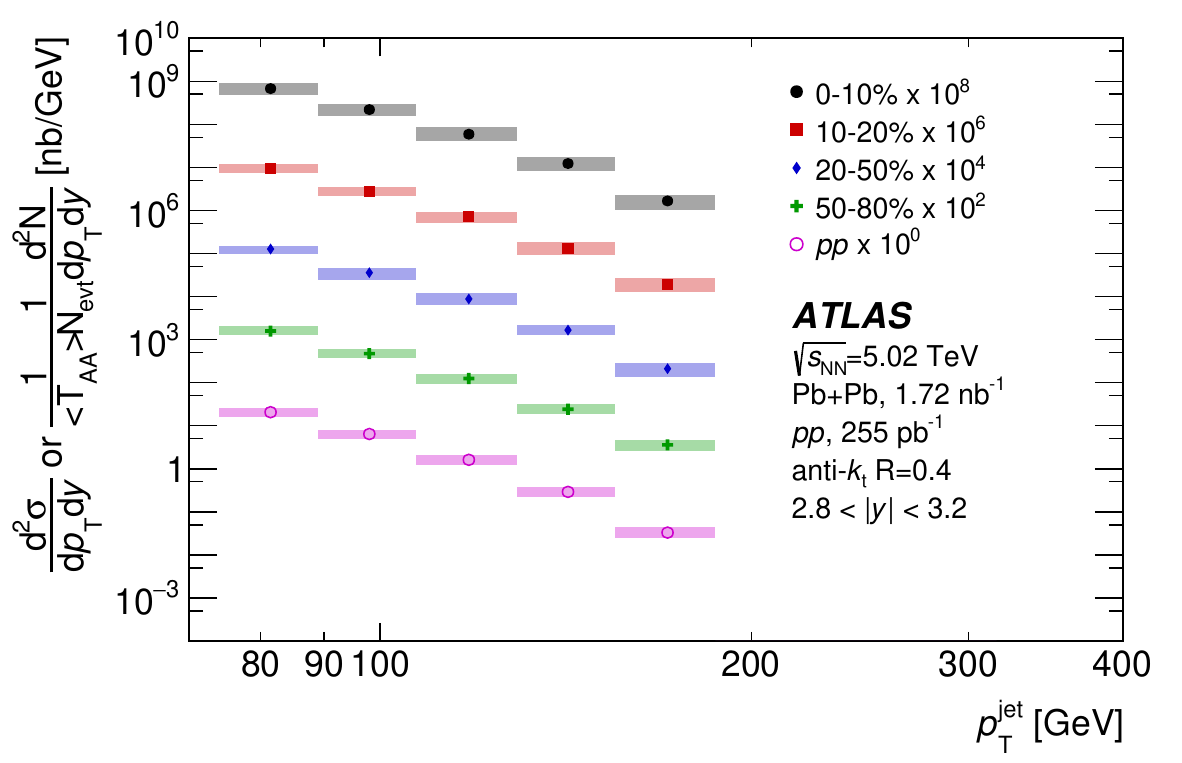}
\includegraphics[width=0.45\linewidth]{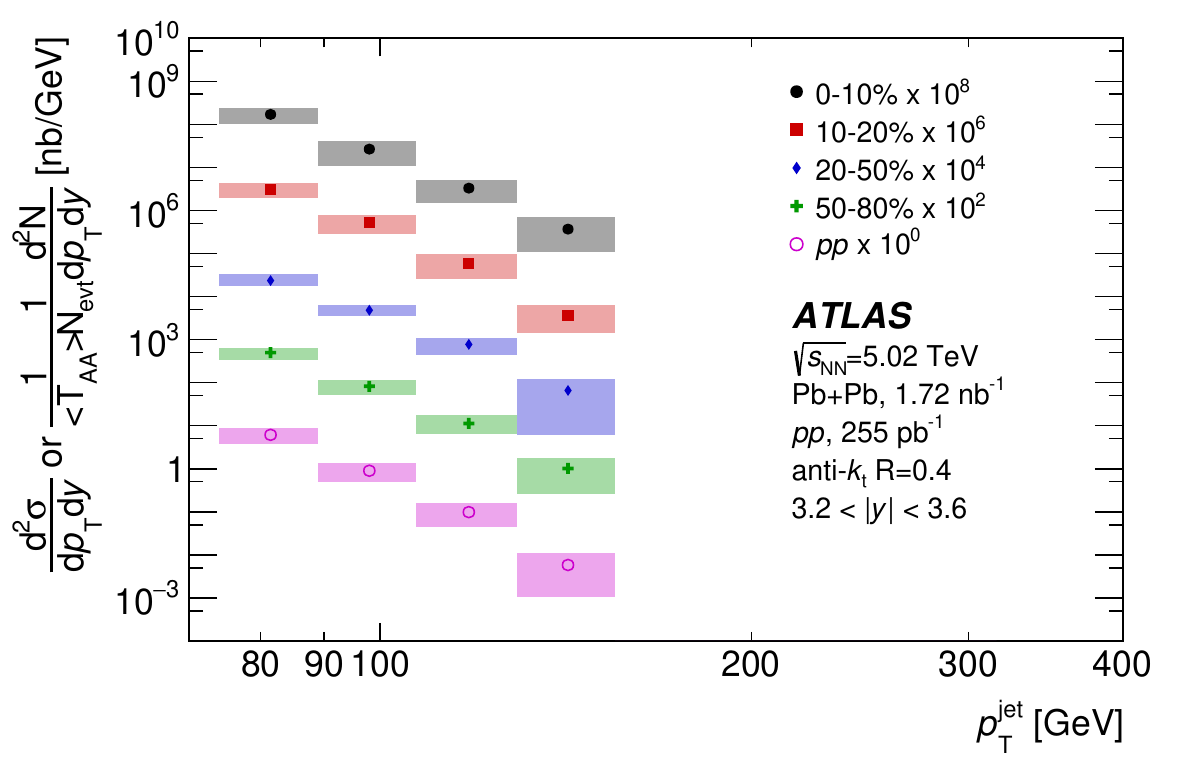}
\caption{$\avgTAA$-normalized jet yields in \PbPb collisions and jet cross-sections in \pp collisions measured in rapidity regions of $2.8<|y|<3.2$ (left) and $3.2<|y|<3.6$ (right). Statistical uncertainties and systematic uncertainties are shown in error bars and shaded boxes around the nominal value, respectively. }
\label{fig:unfolded_jet_yields}
\end{figure}
Figure \ref{fig:unfolded_jet_yields} shows the $\avgTAA$-normalized jet yields in \PbPb collisions and the jet cross-section in \pp collisions. The yields and cross-section are measured as a function of jet $\pt$ in two rapidity regions, $2.8<|y|<3.2$ and $3.2<|y|<3.6$. The jet yields in \PbPb collisions are reported for four centrality intervals. %
The systematic uncertainties shown in the figure also include uncertainties due to luminosity and $\avgTAA$, which are correlated across all data points.

\begin{figure}
\centering
\includegraphics[width=0.45\linewidth]{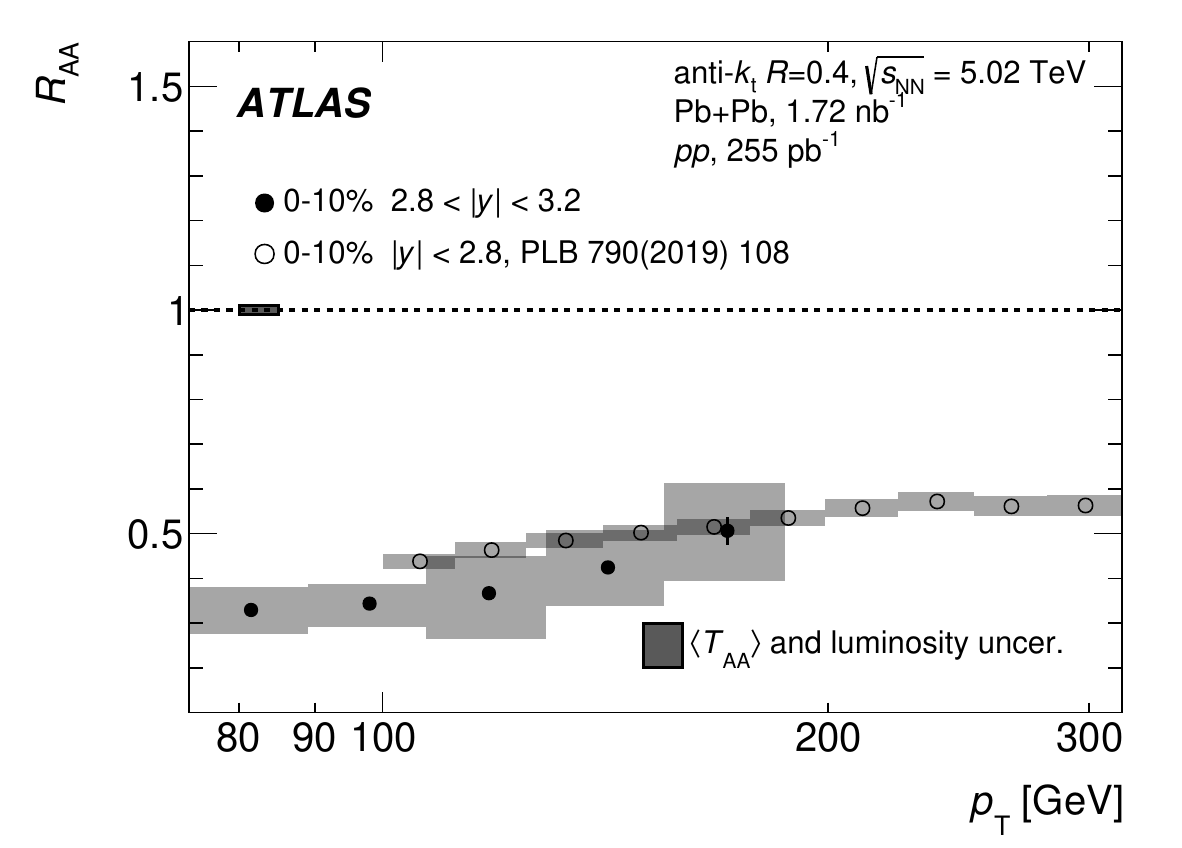}
\includegraphics[width=0.45\linewidth]{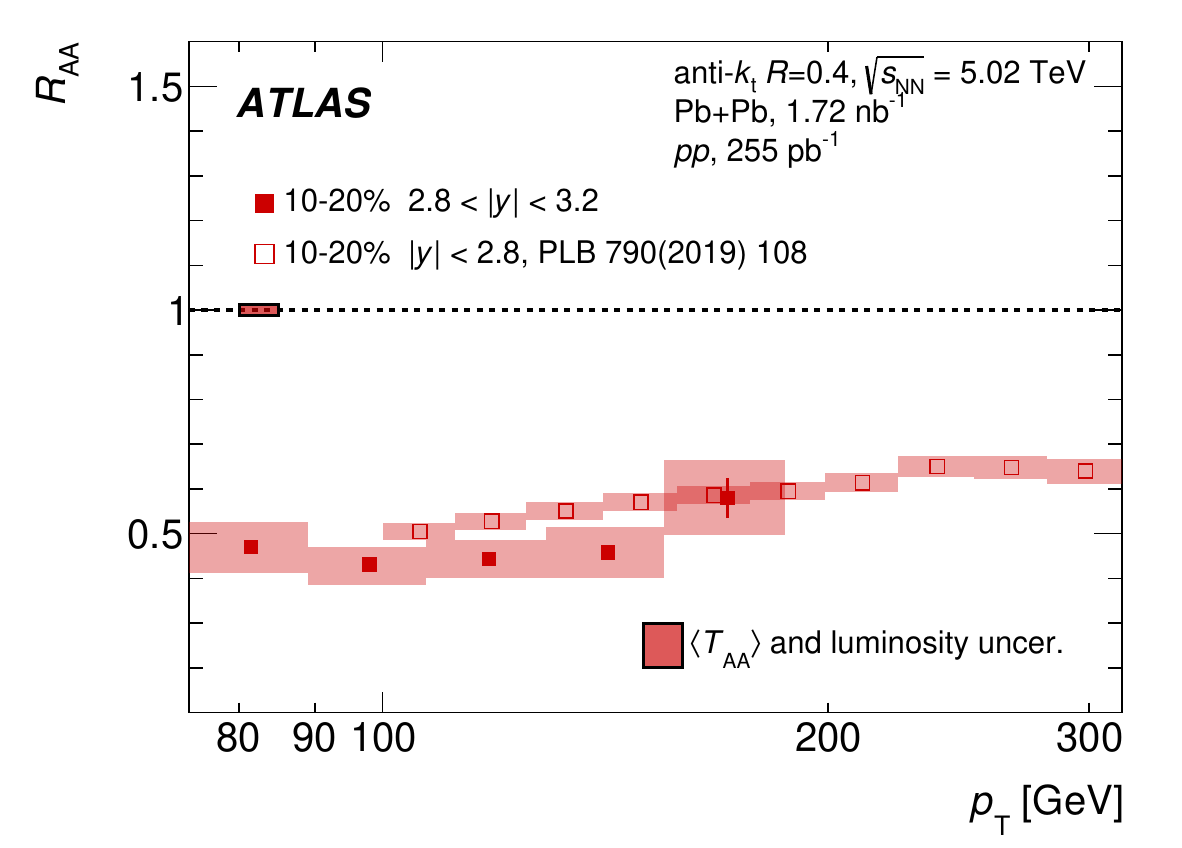}\\
\includegraphics[width=0.45\linewidth]{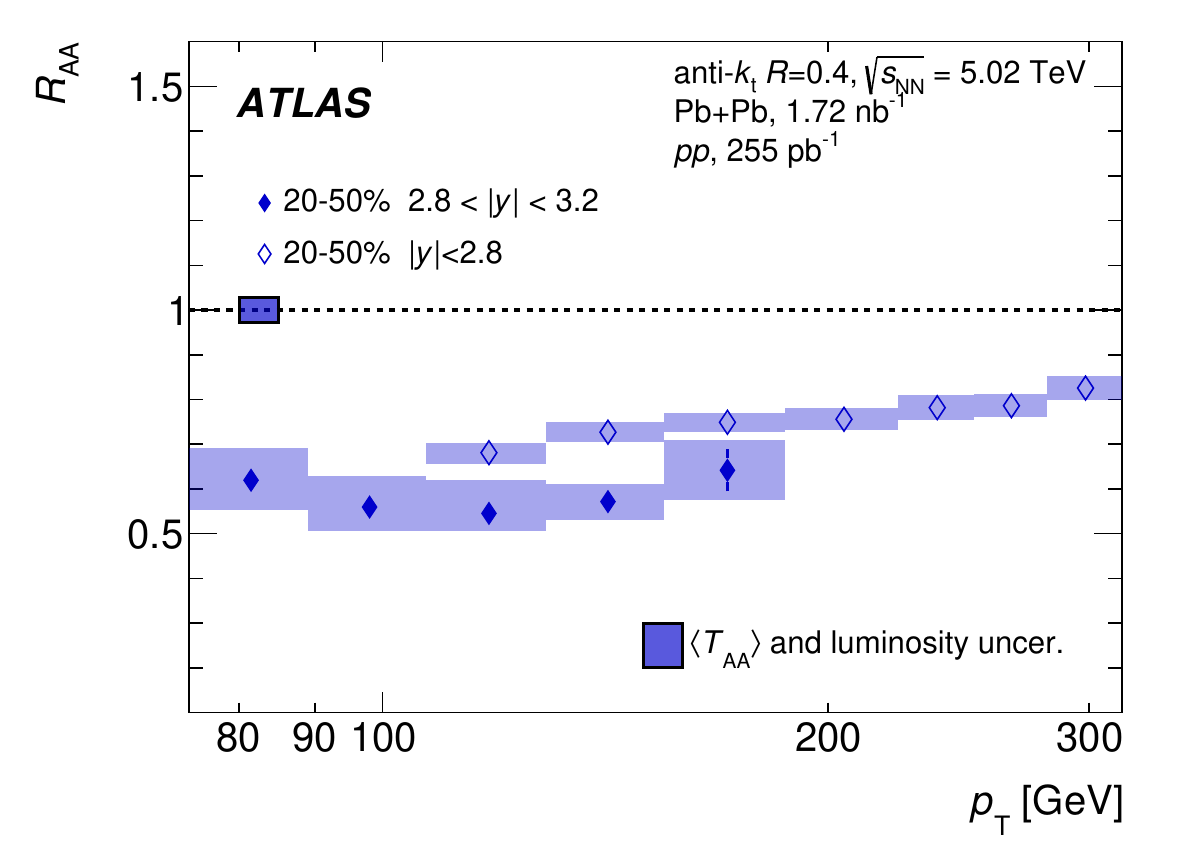}
\includegraphics[width=0.45\linewidth]{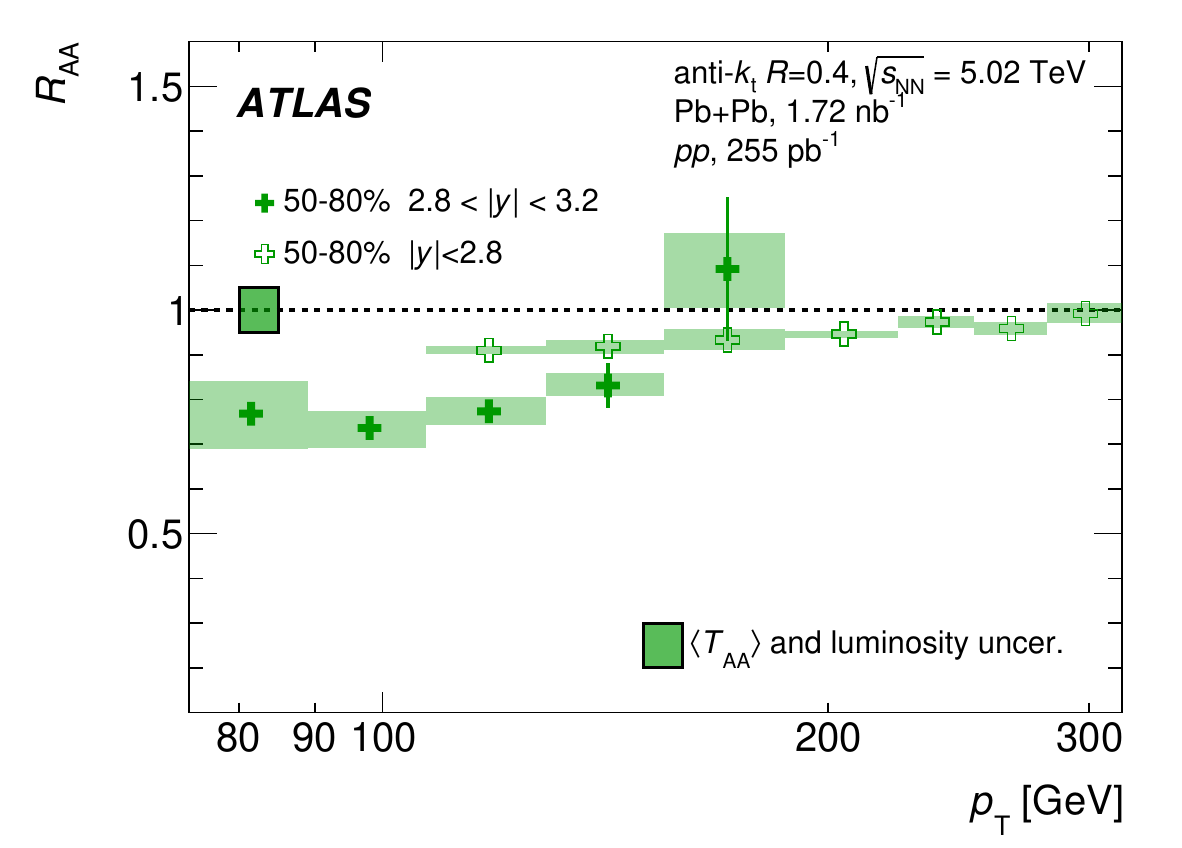}
\caption{Jet $\RAA$ measured in the rapidity interval of \semifwd compared with $\RAA$ measured in the region of $|y|<2.8$.
The four panels show different \PbPb centrality intervals. Bin-wise uncorrelated systematic uncertainties and statistical uncertainties are shown in shaded boxes around the nominal value and error bars, respectively. The fully bin-wise correlated systematic uncertainties are shown as colored boxes at unity representing the quadrature sum of the uncertainties on $\avgTAA$ and $pp$ luminosity.
The results in the $|y|<2.8$ region for the 0--10\% and 10--20\% centrality intervals are taken from Ref.~\cite{HION-2017-10}.
}
\label{fig:unfolded_RAA_semifwd}
\end{figure}
\begin{figure}
\centering
\includegraphics[width=0.45\linewidth]{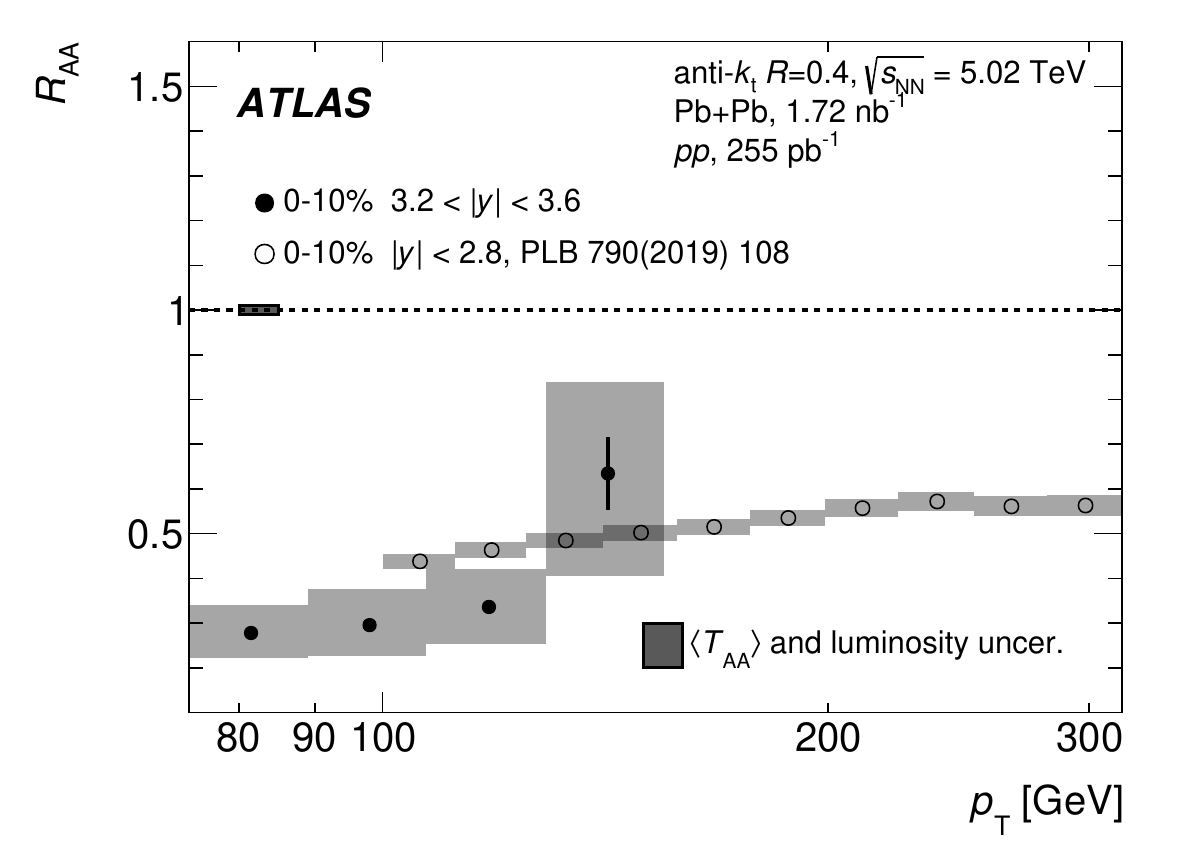}
\includegraphics[width=0.45\linewidth]{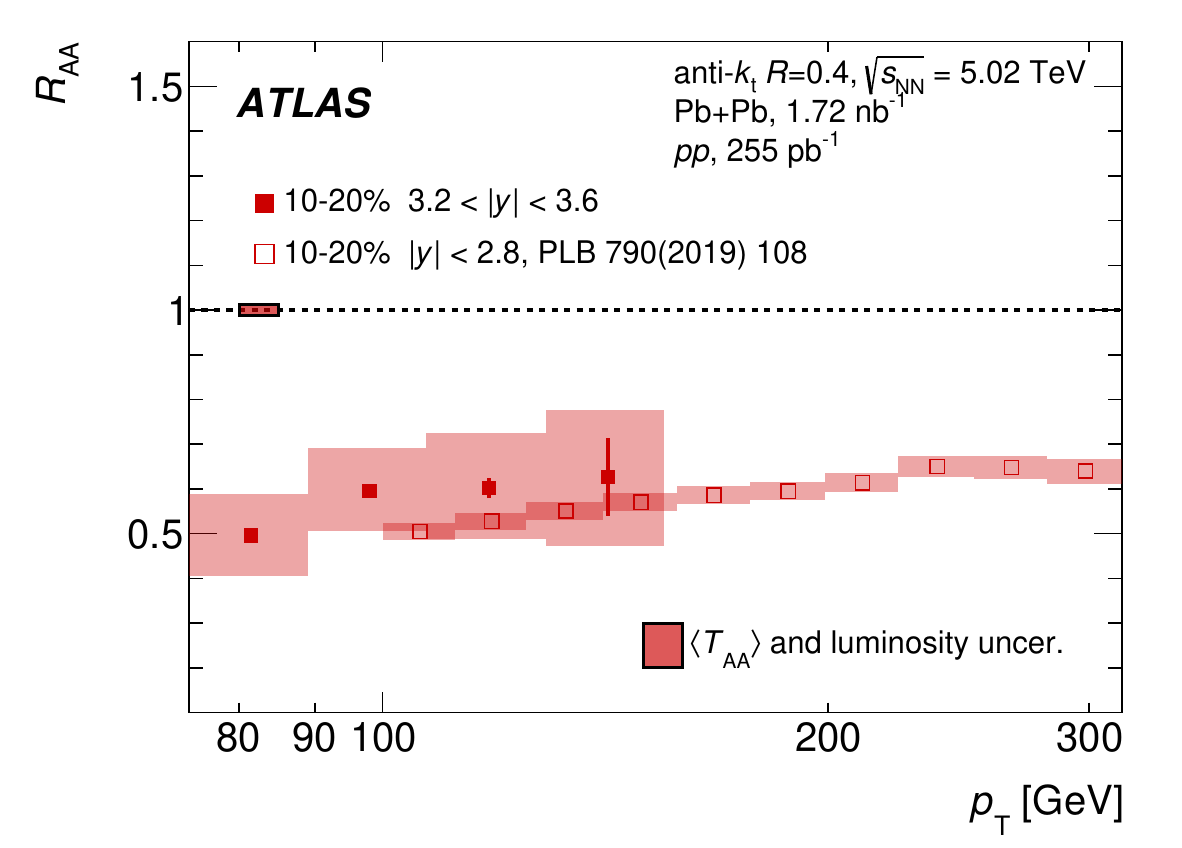}\\
\includegraphics[width=0.45\linewidth]{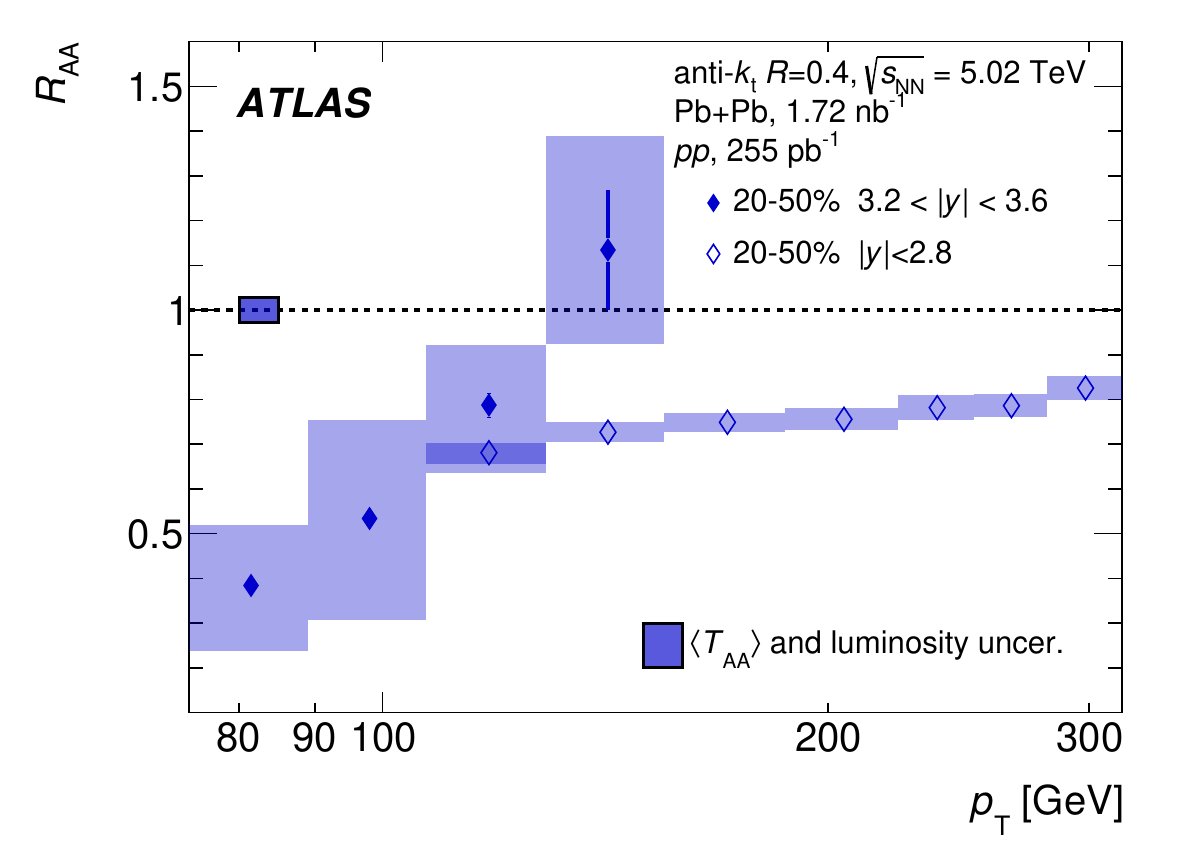}
\includegraphics[width=0.45\linewidth]{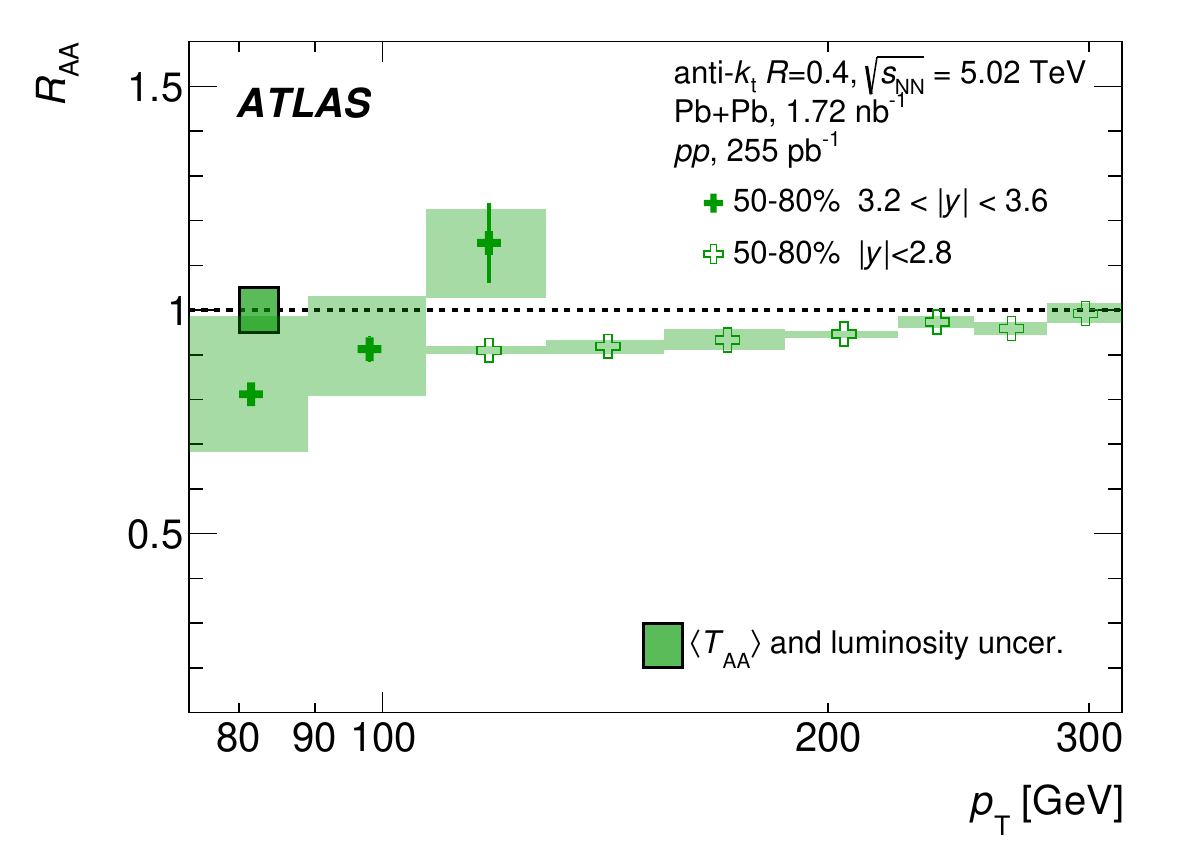}
\caption{Jet $\RAA$ measured in the rapidity interval of \semifwd compared with $\RAA$ measured in the region of $|y|<2.8$. The four panels show different \PbPb centrality intervals. Bin-wise uncorrelated systematic uncertainties and statistical uncertainties are shown in shaded boxes around the nominal value and error bars, respectively. The fully bin-wise correlated systematic uncertainties are shown as colored boxes at unity representing the quadrature sum of the uncertainties on $\avgTAA$ and $pp$ luminosity.
The results in the $|y|<2.8$ region for the 0--10\% and 10--20\% centrality intervals are taken from Ref.~\cite{HION-2017-10}.
}
\label{fig:unfolded_RAA_fwd}
\end{figure}

The measurement of $\RAA$, calculated as the ratio of the $\avgTAA$-normalized per-event jet yield in \PbPb collisions to the jet cross-section in \pp collisions, is shown in Figures~\ref{fig:unfolded_RAA_semifwd} and \ref{fig:unfolded_RAA_fwd} for the rapidity regions $2.8<|y|<3.2$ and $3.2<|y|<3.6$, respectively.
The $\RAA$ is evaluated as a function of jet \pt and compared to the \RAA measured in the rapidity region of $|y|<2.8$.
The results presented in the $|y|<2.8$ region are taken directly from Ref.~\cite{HION-2017-10} for the comparisons in the 0--10\% and 10--20\% centrality intervals, while the results in the 20--50\% and 50--80\% intervals are obtained from a reanalysis of the measurements published in Ref.~\cite{HION-2017-10} to accommodate the centrality binning used in the present analysis. This reanalysis follows the same methods and procedures as those described in Ref.~\cite{HION-2017-10} and gives results consistent with the published measurements in the common centrality and kinematic intervals.

\begin{figure}
\centering
\begin{subfigure}{0.45\textwidth}
\includegraphics[width=\linewidth]{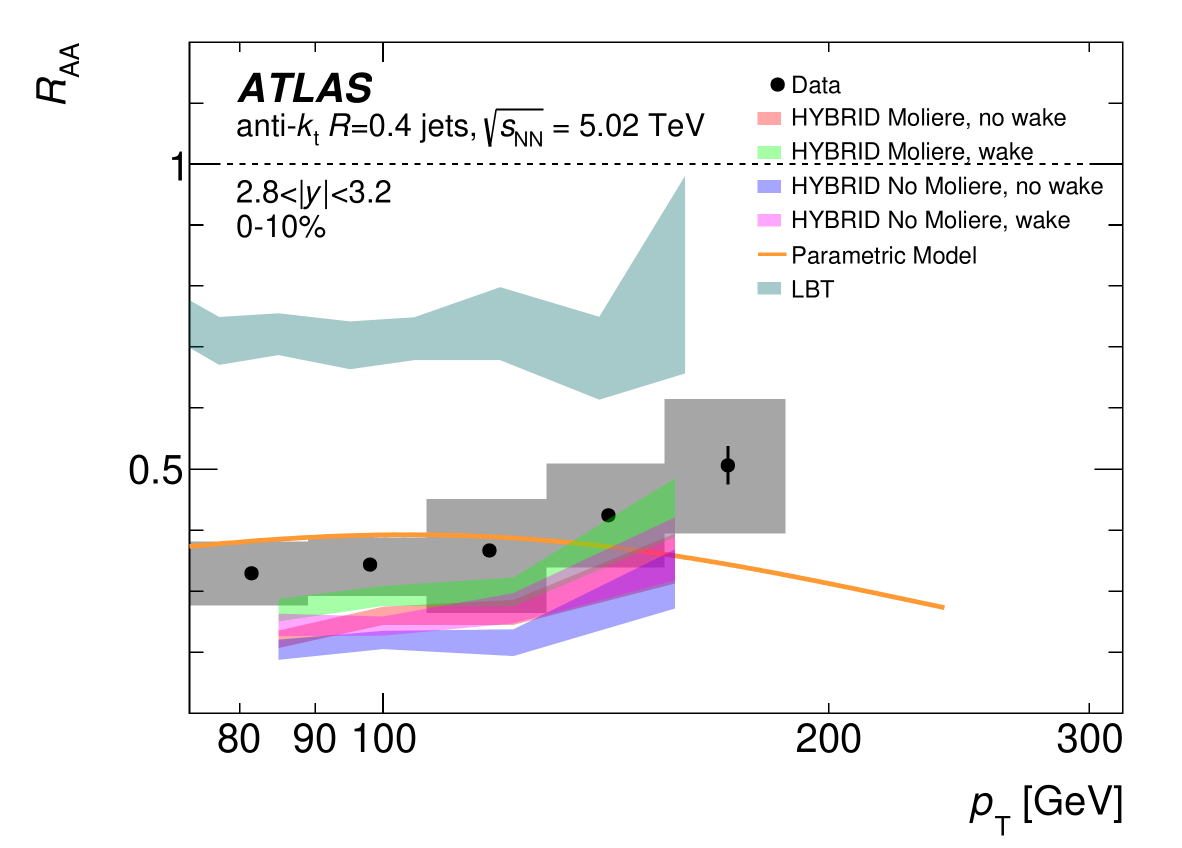}
\caption{} %
\end{subfigure}
\begin{subfigure}{0.45\textwidth}
\includegraphics[width=\linewidth]{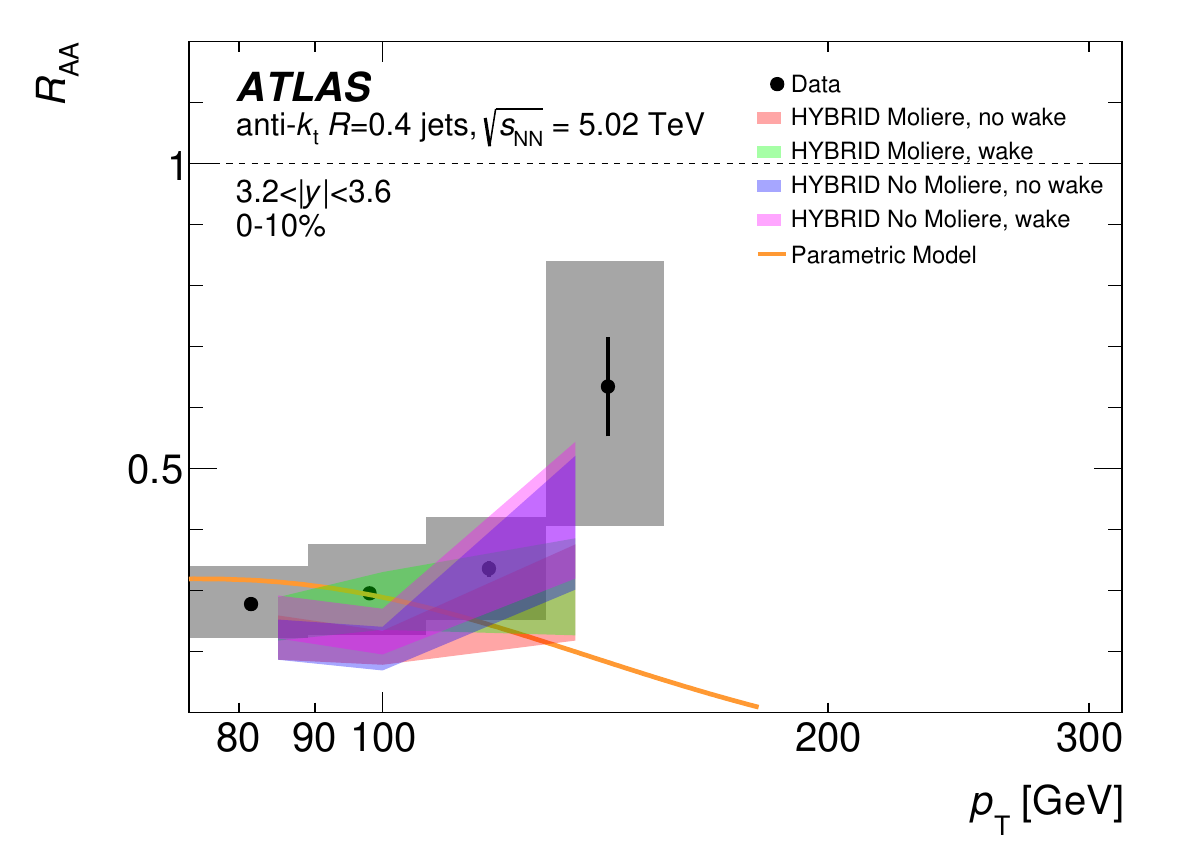}
\caption{} %
\end{subfigure}
\caption{
The $\RAA$ values as a function of jet \pt for the 0--10\% centrality interval measured in (a) $\semifwd$ and (b) $\fwd$ rapidity regions compared with theory predictions. Bin-wise uncorrelated systematic uncertainties and statistical uncertainties are shown in shaded boxes around the nominal value and error bars, respectively.
}
\label{fig:theory_comparison}
\end{figure}

In general, the reported jet $\RAA$ in both rapidity regions increases with increasing jet $\pt$ and gradually increases toward more peripheral \PbPb collisions. While the overall magnitude of $\RAA$ increases with centrality as expected, its $\pt$ dependence does not change significantly with centrality, which is a feature seen in previous studies of this quantity in the $|y|<2.8$ region.
In the $2.8 < |y| < 3.2$ region, the magnitude of jet $\RAA$ shows a tendency to be lower than that measured in the $|y|<2.8$ region, consistent with a trend already suggested by the mid-rapidity $\RAA$ measurement~\cite{HION-2017-10}.
In the $3.2 < |y| < 3.6$ region, the precision of the forward jet measurement does not allow a conclusive statement about possible differences relative to the $|y|<2.8$ region. The interpretation of these trends requires theoretical calculations that account for the rapidity dependence of the initial jet spectra, the quark- and gluon-initiated jet fractions, and the medium evolution. A comparison of the measured $\RAA$ values with such predictions is therefore provided in Figure~\ref{fig:theory_comparison}.

The $\RAA$ values evaluated as a function of jet $\pt$ in the 0--10\% centrality interval are compared with three sets of theoretical predictions for which calculations are available in the forward rapidity region: the Linear Boltzmann Transport model (LBT)~\cite{He:2015pra,He:2018xjv}, four calculations from the Hybrid model~\cite{Casalderrey-Solana:2014bpa,Pablos:2022mrx}, and the Parametric Model~\cite{Spousta:2015fca,Ogrodnik:2024qug}. The LBT model combines a kinetic description of parton propagation with a hydrodynamic description of the underlying medium evolution, while keeping track of thermal recoil partons from each scattering and their further propagation in the medium~\cite{He:2015pra}. The Hybrid model treats the initial splitting processes that form the parton shower as a weak-coupling phenomenon, while modeling the energy lost by these partons to the dense medium as a strongly coupled process~\cite{Casalderrey-Solana:2014bpa}. The Parametric Model infers a generic parameterization of the $\pt$ dependence of energy loss from the data separately for quark-initiated and gluon-initiated jets, and applies the same amount of energy loss at different rapidities, where the flavor composition of jets is different.
Figure~\ref{fig:theory_comparison} shows that the LBT calculations reproduce the shape of $\RAA$ but underpredict the suppression.
The Parametric Model overpredicts the suppression at high-$\pt$,
while the best agreement between data and theory is achieved by the Hybrid model when both Moli{\`e}re scatterings~\cite{DEramo:2018eoy} and the medium wake, describing the response of the QGP to the deposited energy and momentum~\cite{Casalderrey-Solana:2004fdk}, are included.

%
%
%


%

\section{Summary}
\label{sec:summary}

Measurements of the jet cross-section in \pp collisions, the centrality-dependent jet yields in \PbPb collisions, and the jet $\RAA$ are %
performed using data collected by the ATLAS detector at the LHC at a nucleon--nucleon center-of-mass energy of 5.02~\TeV, with a total \PbPb luminosity of 1.72 nb$^{-1}$
and a total \pp luminosity of 255 pb$^{-1}$.

Anti-$k_t$ jets with $R = 0.4$ and $\pt > 74$~\GeV are measured in two rapidity intervals, $2.8 < |y| < 3.2$ and $3.2 < |y| < 3.6$. The unfolded jet yields measured in \PbPb collisions, scaled by the nuclear overlap function, $\avgTAA$, are observed to be suppressed relative to the jet cross-section measured in \pp collisions. The magnitude of \RAA monotonically increases from central to peripheral collisions and with increasing jet $\pt$, while its $\pt$ dependence remains comparable across the measured centrality and rapidity intervals. In the $2.8 < |y| < 3.2$ region, the magnitude of jet $\RAA$ shows a tendency to be lower than that measured in the $|y|<2.8$ region.
In the $3.2<|y|<3.6$ region, the precision of the measurement does not allow a conclusive comparison with the $|y|<2.8$ region.

The results presented in this \paper constitute the first measurement in heavy-ion collisions of jet suppression in the forward-rapidity region.
Compared with mid-rapidity, the forward region is characterized by different admixtures of quark- and gluon-initiated jets, potentially different effective medium densities in the longitudinally expanding QGP, and different levels of selection bias. Measurements in this kinematic regime therefore provide new constraints on theoretical descriptions of jet quenching and the flavor, geometric, and medium-density dependence of parton energy loss.

%

%

%

%


%

%
%
%
%
%
\section*{Acknowledgments}
%

%

%
%

%
%

We thank CERN for the very successful operation of the LHC and its injectors, as well as the support staff at
CERN and at our institutions worldwide without whom ATLAS could not be operated efficiently.

The crucial computing support from all WLCG partners is acknowledged gratefully, in particular from CERN, the ATLAS Tier-1 facilities at TRIUMF/SFU (Canada), NDGF (Denmark, Norway, Sweden), CC-IN2P3 (France), KIT/GridKA (Germany), INFN-CNAF (Italy), NL-T1 (Netherlands), PIC (Spain), RAL (UK) and BNL (USA), the Tier-2 facilities worldwide and large non-WLCG resource providers. Major contributors of computing resources are listed in Ref.~\cite{ATL-SOFT-PUB-2026-001}.

We gratefully acknowledge the support of ANPCyT, Argentina; YerPhI, Armenia; ARC, Australia; BMWFW and FWF, Austria; ANAS, Azerbaijan; CNPq and FAPESP, Brazil; NSERC, NRC and CFI, Canada; CERN; ANID, Chile; CAS, MOST, NSFC and FRFCU, China; Minciencias, Colombia; MEYS CR, Czech Republic; DNRF and DNSRC, Denmark; IN2P3-CNRS and CEA-DRF/IRFU, France; SRNSFG, Georgia; BMFTR, HGF and MPG, Germany; GSRI, Greece; RGC and Hong Kong SAR, China; ICHEP and Academy of Sciences and Humanities, Israel; INFN, Italy; MEXT and JSPS, Japan; CNRST, Morocco; NWO, Netherlands; RCN, Norway; MNiSW, Poland; FCT, Portugal; MNE/IFA, Romania; MSTDI, Serbia; MSSR, Slovakia; ARIS and MVZI, Slovenia; DSI/NRF, South Africa; MICIU/AEI, Spain; SRC and Wallenberg Foundation, Sweden; SERI, SNSF and Cantons of Bern and Geneva, Switzerland; NSTC, Taipei; TENMAK, T\"urkiye; STFC/UKRI, United Kingdom; DOE and NSF, United States of America.

Individual groups and members have received support from BCKDF, CANARIE, CRC and DRAC, Canada; CERN-CZ, FORTE and PRIMUS, Czech Republic; COST, ERC, ERDF, Horizon 2020 and Marie Sk{\l}odowska-Curie Actions, European Union; Investissements d'Avenir Labex, Investissements d'Avenir Idex and ANR, France; DFG and AvH Foundation, Germany; Herakleitos, Thales and Aristeia programmes co-financed by EU-ESF and the Greek NSRF, Greece; BSF-NSF and MINERVA, Israel; NCN and NAWA, Poland; La Caixa Banking Foundation, CERCA and AGAUR programs from Generalitat de Catalunya and PROMETEO and GenT Programmes Generalitat Valenciana, Spain; G\"{o}ran Gustafssons Stiftelse, Sweden; The Royal Society and Leverhulme Trust, United Kingdom; Eric and Wendy Schmidt Fund for Strategic Innovation, United States of America.

In addition, individual members wish to acknowledge support from Chile: Agencia Nacional de Investigaci\'on y Desarrollo (ANID FONDECYT reg. 1230987, FONDECYT 1230812, FONDECYT 1240864, Fondecyt 3240661, Fondecyt Regular 1240721); China: Fundamental Research Funds for the Central Universities (010-63263105), Chinese Ministry of Science and Technology (MOST-2023YFA1605700, MOST-2023YFA1609300), National Natural Science Foundation of China (NSFC 12275265, NSFC-W2543005); Czech Republic: Czech Science Foundation (GACR - 24-11373S), Ministry of Education Youth and Sports (ERC-CZ-LL2327, FORTE CZ.02.01.01/00/22\_008/0004632); EU: H2020 European Research Council (ERC - 101002463); European Union: European Research Council (BARD No. 101116429, ERC - 101219398, ERC 101089007), European Regional Development Fund (HE COFUND GA No.101081355, ERDF), Marie Sklodowska-Curie Actions (GAP-101168829); France: Agence Nationale de la Recherche (ANR-22-EDIR-0002, ANR-24-CE31-0504-01); Germany: Deutsche Forschungsgemeinschaft (DFG - 469666862); China: Research Grants Council (GRF); Italy: Istituto Nazionale di Fisica Nucleare (LHC-MIUR - 28003/2025), Ministero dell'Università e della Ricerca (NextGenEU I53D23000820006 M4C2.1.1, SOE2024\_0000023); Japan: Japan Society for the Promotion of Science (JSPS KAKENHI  JP25H0063, JSPS KAKENHI JP22H04944, JSPS KAKENHI JP24K23939, JSPS KAKENHI JP24KK0251, JSPS KAKENHI JP25H00650, JSPS KAKENHI JP25H01291, JSPS KAKENHI JP25K01011, JSPS KAKENHI JP25K01023, JSPS KAKENHI JP25KK0047); Poland: Polish National Science Centre (NCN 2021/42/E/ST2/00350, NCN OPUS 2023/51/B/ST2/02507, NCN OPUS nr 2022/47/B/ST2/03059, NCN UMO-2019/34/E/ST2/00393, UMO-2022/47/O/ST2/00148, UMO-2023/49/B/ST2/04085, UMO-2023/51/B/ST2/00920, UMO-2024/53/N/ST2/00869); Spain: Agència de Gestió d'Ajuts Universitaris i de Recerca. (AGAUR - 2023 BP 00141), Ministry of Science and Innovation (RYC2019-028510-I, RYC2020-030254-I, RYC2021-031273-I, RYC2022-038164-I), Ministerio de Ciencia, Innovación y Universidades/Agencia Estatal de Investigaci\'on (EU NextGenerationEU (PRTR-C17.I1), PID2022-142604OB-C22); Sweden: Carl Trygger Foundation (Carl Trygger Foundation CTS 22:2312), Swedish Research Council (Swedish Research Council 2023-04654, VR 2021-03651, VR 2022-03845, VR 2022-04683, VR 2023-03403, VR 2024-05451, VR 2025-05940), Knut and Alice Wallenberg Foundation (KAW 2023.0366); United Kingdom: The Binks Trust, Royal Society (NIF-R1-231091); United States of America: U.S. Department of Energy (ECA DE-AC02-76SF00515), John Templeton Foundation (John Templeton Foundation 63206), Neubauer Family Foundation.

%
%


%
\clearpage
\printbibliography
\clearpage
 
\begin{flushleft}
\hypersetup{urlcolor=black}
{\Large The ATLAS Collaboration}

\bigskip

\AtlasOrcid[0000-0002-6665-4934]{G.~Aad}$^\textrm{\scriptsize 102}$,
\AtlasOrcid[0000-0001-7616-1554]{E.~Aakvaag}$^\textrm{\scriptsize 17}$,
\AtlasOrcid[0000-0002-5888-2734]{B.~Abbott}$^\textrm{\scriptsize 121}$,
\AtlasOrcid[0000-0002-0287-5869]{S.~Abdelhameed}$^\textrm{\scriptsize 83b}$,
\AtlasOrcid[0000-0002-1002-1652]{K.~Abeling}$^\textrm{\scriptsize 54}$,
\AtlasOrcid[0000-0001-5763-2760]{N.J.~Abicht}$^\textrm{\scriptsize 48}$,
\AtlasOrcid[0000-0002-8496-9294]{S.H.~Abidi}$^\textrm{\scriptsize 30}$,
\AtlasOrcid[0009-0003-6578-220X]{M.~Aboelela}$^\textrm{\scriptsize 44}$,
\AtlasOrcid[0000-0002-9987-2292]{A.~Aboulhorma}$^\textrm{\scriptsize 36e}$,
\AtlasOrcid[0000-0001-5329-6640]{H.~Abramowicz}$^\textrm{\scriptsize 154}$
% ...
\bigskip
\\

% ...
$^{17}$Department for Physics and Technology, University of Bergen, Bergen; Norway.\\
% ...
$^{30}$Physics Department, Brookhaven National Laboratory, Upton NY; United States of America.\\
% ...
$^{36}$$^{(a)}$Facult\'e des Sciences Ain Chock, Universit\'e Hassan II de Casablanca;$^{(b)}$Facult\'{e} des Sciences, Universit\'{e} Ibn-Tofail, K\'{e}nitra;$^{(c)}$Facult\'e des Sciences Semlalia, Universit\'e Cadi Ayyad, LPHEA-Marrakech;$^{(d)}$LPMR, Facult\'e des Sciences, Universit\'e Mohamed Premier, Oujda;$^{(e)}$Facult\'e des sciences, Universit\'e Mohammed V, Rabat;$^{(f)}$Institute of Applied Physics, Mohammed VI Polytechnic University, Ben Guerir; Morocco.\\
% ...
$^{44}$Physics Department, Southern Methodist University, Dallas TX; United States of America.\\
% ...
$^{48}$Fakult\"{a}t Physik, Technische Universit{\"a}t Dortmund, Dortmund; Germany.\\
% ...
$^{54}$II. Physikalisches Institut, Georg-August-Universit\"{a}t G\"ottingen, G\"ottingen; Germany.\\
% ...
$^{83}$$^{(a)}$Khalifa University of Science and Technology, Abu Dhabi;$^{(b)}$New York University Abu Dhabi, Abu Dhabi;$^{(c)}$United Arab Emirates University, Al Ain;$^{(d)}$University of Sharjah, Sharjah; United Arab Emirates.\\
% ...
$^{102}$CPPM, Aix-Marseille Universit\'e, CNRS/IN2P3, Marseille; France.\\
% ...
$^{121}$Homer L. Dodge Department of Physics and Astronomy, University of Oklahoma, Norman OK; United States of America.\\
% ...
$^{154}$Raymond and Beverly Sackler School of Physics and Astronomy, Tel Aviv University, Tel Aviv; Israel.\\
% ...
$^{a}$ Also at Affiliated with an institute formerly covered by a cooperation agreement with CERN.\\
$^{b}$ Also at An-Najah National University, Nablus; Palestine.\\
$^{c}$ Also at Borough of Manhattan Community College, City University of New York, New York NY; United States of America.\\
$^{d}$ Also at Center for Interdisciplinary Research and Innovation (CIRI-AUTH), Thessaloniki; Greece.\\
$^{e}$ Also at Centre of Physics of the Universities of Minho and Porto (CF-UM-UP); Portugal.\\
$^{f}$ Also at CERN, Geneva; Switzerland.\\
% ...

\end{flushleft}


%

%
%

%

%
%
%
%

\end{document}